\documentclass[10pt,fullpage]{article}

\usepackage{fullpage}
\usepackage{graphicx}
\usepackage{amsmath}
\usepackage{amssymb}
\usepackage{amstext}
\usepackage{amsthm}
\usepackage{epsfig}
\usepackage{placeins}
\usepackage{subfig}
\usepackage{paralist}
\usepackage{caption}
\usepackage{multirow}
\usepackage{algorithm}
\usepackage{algorithmic}
\usepackage{comment}
\usepackage[obeyspaces,spaces]{url}
\usepackage{balance}
\usepackage{enumitem}
\usepackage{hyperref}
\usepackage{times}

\newcommand{\eat}[1]{}

\makeatletter
\newenvironment{sql}%
 {\vskip 5pt\begin{list}{}{%
  \setlength{\topsep}{0pt}\setlength{\partopsep}{0pt}\setlength{\parskip}{0pt}%
  \setlength{\parsep}{0pt}\setlength{\labelwidth}{0pt}%
  \setlength{\rightmargin}{0pt}\setlength{\leftmargin}{0pt}%
  \setlength{\labelsep}{0pt}%
  \obeylines\@vobeyspaces\normalfont\ttfamily%
  \item[]}}
 {\end{list}\vskip5pt\noindent}
\makeatother

\begin{document}

\date{February 2018}

\title{Stochastic Gradient Descent on Highly-Parallel Architectures}

\author{
Yujing Ma \hspace*{2cm} Florin Rusu \hspace*{2cm} Martin Torres\\
{University of California Merced}\\
{\texttt{\{yma33, frusu, mtorres58\}@ucmerced.edu}}
}

\maketitle

%%%%%%%%%%%%%%%%%%%%%%%%%%%%%%%%%%%%%%%%%%%%%%%%%%%%%%%%%%%%%%%%%
%\input{abstract}

\begin{abstract}
There is an increased interest in building data analytics frameworks with advanced algebraic capabilities both in industry and academia. Many of these frameworks, e.g., TensorFlow and BIDMach, implement their compute-intensive primitives in two flavors---as multi-thread routines for multi-core CPUs and as highly-parallel kernels executed on GPU. Stochastic gradient descent (SGD) is the most popular optimization method for model training implemented extensively on modern data analytics platforms. While the data-intensive properties of SGD are well-known, there is an intense debate on which of the many SGD variants is better in practice. In this paper, we perform a comprehensive study of parallel SGD for training generalized linear models. We consider the impact of three factors -- computing architecture (multi-core CPU or GPU), synchronous or asynchronous model updates, and data sparsity -- on three measures---hardware efficiency, statistical efficiency, and time to convergence. In the process, we design an optimized asynchronous SGD algorithm for GPU that leverages warp shuffling and cache coalescing for data and model access. We draw several interesting findings from our extensive experiments with logistic regression (LR) and support vector machines (SVM) on five real datasets. For synchronous SGD, GPU always outperforms parallel CPU---they both outperform a sequential CPU solution by more than 400X. For asynchronous SGD, parallel CPU is the safest choice while GPU with data replication is better in certain situations. The choice between synchronous GPU and asynchronous CPU depends on the task and the characteristics of the data. As a reference, our best implementation outperforms TensorFlow and BIDMach consistently. We hope that our insights provide a useful guide for applying parallel SGD to generalized linear models.
\end{abstract}

%%%%%%%%%%%%%%%%%%%%%%%%%%%%%%%%%%%%%%%%%%%%%%%%%%%%%%%%%%%%%%%%%
%\input{introduction}

\section{INTRODUCTION}\label{sec:intro}

Stochastic gradient descent (SGD) is the most popular optimization method to train analytics models, e.g., the back-propagation algorithm for deep neural networks~\cite{deep-nets-sgd}, in a wide variety of application domains ranging from image~\cite{cvpr-sgd} and speech~\cite{speech-sgd} recognition to finance~\cite{finance-sgd}. SGD is implemented in a form or another by every modern analytics system, including Google's Brain~\cite{google-brain}, Microsoft's Project Adam~\cite{project-adam} and Vowpal Wabbit~\cite{vowpal-wabbit}, IBM's SystemML~\cite{systemml}, Pivotal's MADlib \cite{madlib}, and Spark's MLlib~\cite{mllib}. Since these billion-dollar enterprises depend on processes which rely on SGD, it is important to understand its optimal behavior and limitations on modern computing architectures.

%%%%%%%%%%%%%%%%%%%%%%%%%%%%%%%%%%%%%%
\textbf{Motivation.}
Over the past decade, CPU design has been moving towards highly-parallel architectures with tens of cores on a die. The culmination of this trend is best exemplified by the current Graphics Processing Units (GPU) having thousands of cores\footnote{\url{https://en.wikipedia.org/wiki/Nvidia_Tesla}}. GPUs are assumed to be the ideal platform for analytics model training due to the compute-intensive nature of the task. This is exemplified by the extensive GPU support across many analytics frameworks, e.g., Caffe\footnote{\url{http://caffe.berkeleyvision.org/}}, TensorFlow\footnote{\url{https://www.tensorflow.org/}}, MXNet\footnote{\url{https://mxnet.incubator.apache.org/}}, BIDMach\footnote{\url{https://github.com/BIDData/BIDMach}}, SINGA~\cite{singa}, Theano\footnote{\url{https://github.com/Theano/Theano}}, and Torch\footnote{\url{http://torch.ch/}}. Published results that compare CPU and GPU implementations, however, do not support the assumption that GPU is always superior~\cite{Intel-GPU-CPU-comp,google-brain,tensorflow}. Quite the opposite, it is often the case that the CPU optimizer outperforms the GPU implementation, even though the degree of parallelism is much lower. A possible reason is the choice of the SGD algorithm. The asynchronous Hogwild-family of algorithms~\cite{hogwild,bismarck,dimm-witted,hogbatch,buckwild,hogwild-disk} are the preferred SGD implementation on multi-core CPUs due to their simplicity -- the parallel code is identical to the serial one, without any synchronization primitives -- and near-linear scaling across a variety of analytics tasks~\cite{RRTB12,LWR+14,DJM13}. The SGD solutions on GPU resort to a synchronous implementation in which only the highly-optimized linear algebra kernels are offloaded to the GPU. The reasons behind this strategy are the original role GPUs had as accelerators for certain classes of computations and the intricate data access pattern incurred by asynchronous execution. The algorithmic difference in model update strategy -- which is an open debate both in theoretical circles~\cite{sync-vs-asynch-sgd,asynch-sgd-delay-comp} and practice~\cite{tensorflow} -- makes a direct comparison between SGD on CPU and GPU challenging. As far as we know, there is no work that performs an in-depth comparison across architectures and SGD algorithms. Given its central role in analytics training, we believe it is imperative to identify which SGD algorithm performs better on which architecture and what type of data.

%%%%%%%%%%%%%%%%%%%%%%%%%%%%%%%%%%%%%%
\textbf{Problem.}
We briefly present the setup for model training using SGD. The input data is a matrix in $\mathbb{R}^{N\times d}$ containing $N$ $d$-dimensional training examples. The goal is to find a $d$-dimensional vector that minimizes the (convex) loss function over the examples specific to each model. SGD makes several complete passes over the input data and updates the model one or several times in each pass. SGD performance is measured by the time it takes to reach the loss function minimum. This depends both on the number of passes over the data and the time per pass. In parallel SGD, the input data is partitioned across threads which share a single common model. While this reduces the time per pass, it has the potential to increase the number of passes. The overall effect depends on several factors---including parallelization strategy, model update, data characteristics, and loss function type.

%%%%%%%%%%%%%%%%%%%%%%%%%%%%%%%%%%%%%%
\textbf{Contributions.}
In this paper, \textit{we perform the first comprehensive study of parallel SGD for training generalized linear models that investigates the combined impact of three axes -- computing architecture, model update strategy, and data sparsity -- on three measures---hardware efficiency, statistical efficiency, and overall time to convergence}. We allocate a significant part of the study to the design of a novel asynchronous SGD algorithm on GPU -- a missing topic in the existing literature -- that explores exhaustively possible data organization, access, and replication alternatives. To this end, \textit{we introduce an optimized asynchronous SGD algorithm for GPU that leverages architectural characteristics such as warp shuffling and cache coalescing for data and model access}.

%%%%%%%%%%%%%%%%%%%%%%%%%%%%%%%%%%%%%
\begin{figure*}[htbp]
\begin{minipage}{.5\textwidth}
\centering
\includegraphics[width=.7\textwidth]{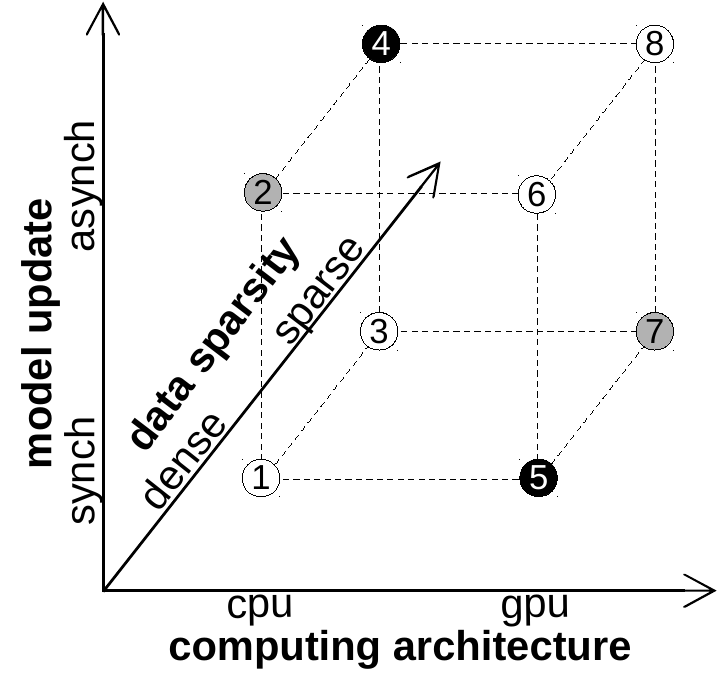}
\caption{Exploratory axes.}
\label{fig:study-axes}
\end{minipage}
\begin{minipage}{.5\textwidth}
\centering
\includegraphics[width=.7\textwidth]{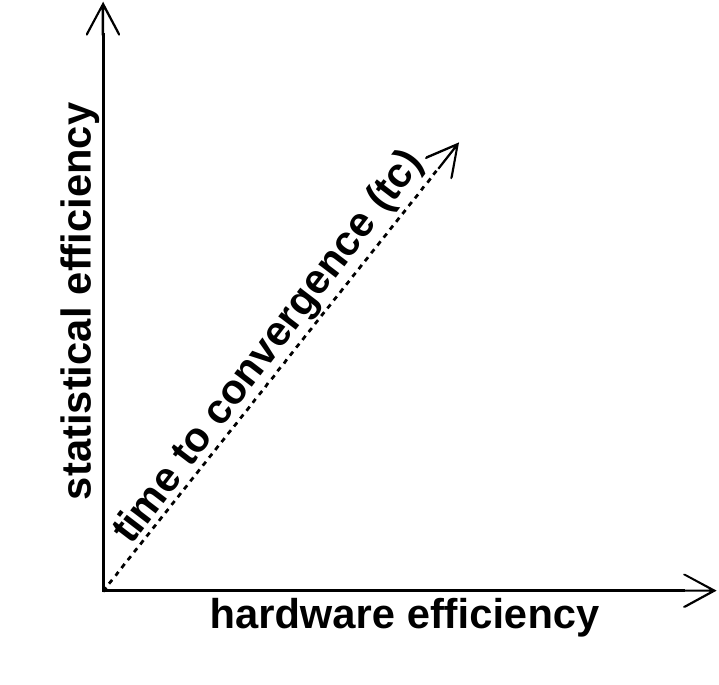}
\caption{Performance axes.}
\label{fig:perf-axes}
\end{minipage}
\end{figure*}
%%%%%%%%%%%%%%%%%%%%%%%%%%%%%%%%%%%%%

%%%%%%%%%%%%%%%%%%%%%%%%%%%%%%%%%%%%%%
\textbf{Exploratory axes.}
Figure~\ref{fig:study-axes} depicts the space of the three axes studied in this work. On the \textit{computing architecture} axis, we consider multi-core CPUs with Non-Uniform Memory Access (NUMA) and many-core GPUs with wide Single Instruction Multiple Data (SIMD) processing units. The specific representatives of the two architectures we use in our work are a dual-socket machine with two 14-core 28-thread Intel Xeon E5-2660 v4 CPUs (56 threads overall, 256 GB memory) and an NVIDIA Tesla K80 GPU with 2496 cores, a 32-wide SIMD unit, and 24 GB memory. Both of them are top-of-the-line representatives in their architectural class. The \textit{model update} strategies we consider are synchronous and asynchronous. Synchronous updates follow a transactional semantics and allow a single thread to update the model. While this strategy limits the range of parallel execution inside the SGD algorithm, it is suitable for batch-oriented high-throughput GPU processing. In the asynchronous strategy, multiple threads update the model concurrently. Our focus is on the Hogwild algorithm which ignores any synchronization to the shared model. \textit{Data sparsity} represents the third axis. At one extreme, we have dense data in which there is a non-zero entry for each feature in every training example. This allows for a complete dense 2-D matrix representation. When the model is large, it is often the case that the examples have only a few non-zero features. A sparse matrix format, e.g., Compressed Sparse Row (CSR), is the only alternative that fits in memory in this case.

\noindent
Out of the eight possible combinations, a limited set is implemented in practice---the full circles in Figure~\ref{fig:study-axes}. The majority of the GPU solutions implement synchronous model updates over dense data, while the CPU implementations use asynchronous Hogwild which is suited for sparse data---the darker circles in the figure. In this paper, we explore the complete space and map the remaining combinations. \textit{We design an efficient Hogwild algorithm for GPU that is carefully tuned to the underlying hardware architecture, the SIMD execution model, and the deep GPU memory hierarchy.} The considerably larger number of threads available on the GPU and their complex memory access pattern pose significant data management challenges in achieving an efficient implementation with optimal convergence. The \textit{Hogwild GPU kernel} considers data access path and replication strategies for data and the model to identify the optimal configuration for a given task-dataset input. Moreover, we introduce specific optimizations to enhance the coalesced memory access for SIMD processing. \textit{Synchronous SGD on CPU} turns out to be a simplification of the GPU solution. Instead of executing the linear algebra kernels on the GPU, invoke functions on the CPU. This is easily achieved by using computational libraries with dual CPU and GPU support, e.g., ViennaCL\footnote{\url{http://viennacl.sourceforge.net/}}. Since the CPU functions do not require data transfer, they have the potential to outperform a sequence of GPU kernels that are not cross-optimized. The massive number of powerful threads in our testbed CPU provides an additional boost in performance.

%%%%%%%%%%%%%%%%%%%%%%%%%%%%%%%%%%%%%%
\textbf{Performance axes.}
Figure~\ref{fig:perf-axes} depicts the three axes across which we measure the performance of the SGD algorithms. The \textit{hardware efficiency} measures the average time to do a complete pass -- or iteration -- over the training examples. Ideally, the larger the number of physical threads, the shorter an iteration takes since each thread has less data to work on---thus, higher hardware efficiency. In practice, though, this holds only when there is no interaction between threads---even then, the size and location of data can be limiting factors. In asynchronous SGD, however, the model is shared by all (or a group of) the threads. This poses a difficult challenge both in the CPU and GPU case. For CPU, the implicit cache coherency mechanism across cores can decrease the hardware efficiency dramatically. Non-coalesced memory accesses inside a SIMD unit have the same effect on GPU. The \textit{statistical efficiency} measures the number of passes over the data until a certain value of the loss function is achieved, e.g., within $1\%$ of the minimum. This number is architecture-independent for synchronous model updates which are executed at the end of each pass. In the case of asynchronous model updates during a data pass, however, the number and order of updates may have a negative impact on the statistical efficiency. The third performance axis is represented by the \textit{time to convergence}. This is, essentially, the product between the hardware and statistical efficiency. The reason we include it as an independent axis is because there are situations when two algorithms have reversed hardware and statistical efficiency -- algorithm A has better hardware efficiency than algorithm B and worse statistical efficiency -- and only the time to convergence allows for a full comparison. Such a case is common for synchronous and asynchronous updates and for CPU and GPU execution, respectively.

%%%%%%%%%%%%%%%%%%%%%%%%%%%%%%%%%%%%%%%%%%%%%
\textbf{Summary of results.}
We organize the results based on the components of the exploratory axes. In the following, we present only the main findings, while we discuss the details in the experimental evaluation (Section~\ref{sec:experiments}):
\begin{compactitem}
\item For synchronous SGD, GPU is always faster than parallel CPU in time per iteration and, thus, in time to convergence. The gap between GPU and parallel CPU is more than 5X on sparse data, while super-linear speedup of more than 400X is achieved over sequential CPU on cached data.
\item The optimized asynchronous Hogwild algorithm we design for GPU uses different data access path + model replication + data replication configurations for different tasks and datasets---identifying the optimal configuration is highly-dependent on all these properties. However, even with these optimizations, asynchronous GPU outperforms (parallel) CPU in time to convergence only in a limited number of situations despite better hardware efficiency---the reason is poor statistical efficiency due to heavy model update conflicts.
\item While GPU is the optimal architecture for synchronous SGD and CPU is optimal for asynchronous SGD, choosing the better of synchronous GPU and asynchronous CPU is task- and dataset-dependent.
\item Our SGD implementations always outperform the synchronous solutions from TensorFlow and BIDMach in time per iteration, number of iterations to convergence, and time to convergence.
\end{compactitem}

%%%%%%%%%%%%%%%%%%%%%%%%%%%%%%%%%%%%%%%%%%%%%
\textbf{Outline.}
In Section~\ref{sec:grad-desc}, we introduce SGD and classify it according to the exploratory axes. The parallel architecture of multi-core CPUs and modern GPUs is presented in Section~\ref{sec:architecture}. The SGD implementation and its optimizations are discussed in Section~\ref{sec:synch-sgd} -- synchronous -- and Section~\ref{sec:asynch-sgd}---asynchronous. Section~\ref{sec:experiments} presents the experimental results. Related work is discussed in Section~\ref{sec:rel-work}, while Section~\ref{sec:conclusions} concludes the paper.

%%%%%%%%%%%%%%%%%%%%%%%%%%%%%%%%%%%%%%%%%%%%%%%%%%%%%%%%%%%%%%%%%
%\input{grad-desc}

\section{STOCHASTIC GRADIENT DESCENT}\label{sec:grad-desc}

Consider the following model training problem with a linearly separable objective function:
\begin{equation}\label{eq:optim-form}
\Lambda(\vec{w}) = \textit{min}_{w \in \mathbb{R}^{d}} \sum_{i=1}^{N} f\left(\vec{w}; \vec{x_{i}}, y_{i}\right)
\end{equation}
in which a $d$-dimensional vector $\vec{w}$, $d \geq 1$, i.e., the model, has to be found such that the objective function is minimized. The constants $\vec{x_{i}}$ and $y_{i}$, $1 \leq i \leq N$, correspond to the feature vector of the $\text{i}^{\text{th}}$ data example and its scalar label, while $f$ is the loss. For example, the loss corresponding to binary classification with logistic regression (LR) and support vector machines (SVM) is $f_{\textit{LR}}(\vec{w}) = \log\left(1+e^{-y_{i} \vec{x}_{i} \cdot \vec{w}}\right)$ and $f_{\textit{SVM}}(\vec{w}) = \textit{max}\left(0, 1 - y_{i} \vec{x}_{i} \cdot \vec{w}\right)$, respectively.

SGD is an iterative optimization algorithm for solving this class of model training problems. It starts from an arbitrary model which is updated iteratively based on a batch of $B$ random training examples $\vec{\vec{X}}_{k}$ and their corresponding scalar labels $\vec{Y}_{k}$. The updated model is computed by moving along the opposite direction of the loss function gradient $\overrightarrow{\nabla f}$. The gradient is a $d$-dimensional vector consisting of entries given by the partial derivative with respect to each dimension, i.e., $\overrightarrow{\nabla f}(\vec{w}) = \left[\frac{\partial f(\vec{w})}{\partial{w_{1}}},\dots,\frac{\partial f(\vec{w})}{\partial{w_{d}}}\right]$. For example, the gradients for LR and SVM are defined as:
\begin{equation*}
\frac{\partial f_{\textit{LR}}(\vec{w})}{\partial{w_{j}}} = x_{ij} \left( -y_{i} \frac {e^{-y_{i} \vec{x}_{i} \cdot \vec{w}}} {1+e^{-y_{i} \vec{x}_{i} \cdot \vec{w}}} \right)
\hspace*{2cm}
\frac{\partial f_{\textit{SVM}}(\vec{w})}{\partial{w_{j}}} =
	\left\{ \begin{array}{rl}
		-y_{i} x_{ij}, & \mbox{if} \hspace*{0.25cm} y_{i} \vec{x}_{i} \cdot \vec{w} < 1 \\
		0, & \mbox{otherwise}
	\end{array} \right.
\end{equation*}

\subsection{Batch and Incremental SGD}\label{ssec:sgd:b-i}

%%%%%%%%%%%%%%%%%%%%%%%%%%%%%%%%%%%%%
\begin{figure*}[h]
\begin{minipage}{.42\textwidth}
The step size $\alpha$, the batch size $B$, and the number of iterations or epochs $t$ are parameters specific to SGD. They are known as hyper-parameters of the model training problem, wherein the dimensions of $\vec{w}$ are typically called the parameters of the model. Depending on the value of $B$, SGD can be classified into incremental ($B=1$), mini-batch ($1 < B < N$), and batch ($B=N$). While mini-batch is standard SGD, incremental and batch correspond to extreme cases of $B$ which allow for an essential data access optimization. Random access to the training examples is replaced by sequential access which is orders of magnitude more efficient---especially for massive data that do not fit in the primary storage. Moreover, they discard the batch size hyper-parameter, thus, simplifying the optimization algorithm. 
\end{minipage}
\hfill
\begin{minipage}{.54\textwidth}
\underline{\textbf{Algorithm 1} Stochastic Gradient Descent (SGD)}
\algsetup{linenodelimiter=.}

\begin{algorithmic}[1]\label{alg:sgd}

\REQUIRE ~~\\
Training examples $\vec{\vec{X}} \in \mathbb{R}^{N\times d}$ and their labels $\vec{Y} \in \mathbb{R}^{N}$\\
Loss function $f$ and its gradient $\overrightarrow{\nabla f}$\\
Initial model $\vec{w} \in \mathbb{R}^{d}$ and step size $\alpha \in \mathbb{R}$\\
Number of epochs $t$ and batch size $B$

\FOR {$k=1$ \textbf{to} $t$}
	\item[\hspace*{.9cm}\textbf{\underline{OPTIMIZATION EPOCH}}]
  \STATE Select a random subset $\vec{\vec{X}}_{k} = \{ \vec{x}_{i_{1}},\dots,\vec{x}_{i_{B}} \}$ of $B$\\
  examples and their labels $\vec{Y}_{k} = \{ y_{i_{1}},\dots,y_{i_{B}} \}$
  \STATE Compute gradient estimate: $\vec{g} \leftarrow \sum_{\vec{\vec{X}}_{k}, \vec{Y}_{k}} {\overrightarrow{\nabla f} \left(\vec{w}\right)}$
  \STATE Update model: $\vec{w} \leftarrow \vec{w} - \alpha \vec{g}$\label{alg:line:model-depend}
\ENDFOR

\RETURN $\vec{w}$

\end{algorithmic}
\hfill
\end{minipage}
\end{figure*}
%%%%%%%%%%%%%%%%%%%%%%%%%%%%%%%%%%%%%

\noindent
The optimization epoch in the SGD algorithm becomes a linear scan over the training dataset in which the gradient is incrementally computed (batch SGD) and the model updated (incremental SGD). The differences between the two algorithms are in how the gradient is computed (estimated) and how many times the model is updated. In batch SGD, the gradient is computed exactly using all the $N$ examples in the training dataset and the model is updated only once per epoch. Incremental SGD is at the other extreme. The gradient is approximated using a single example and the model is also updated for every example---$N$ times per epoch. While identical from a computational perspective, the two algorithms are significantly different in terms of convergence. It is a well-known fact that incremental SGD has a convergence rate as much as $N$ times faster than batch SGD for large $N$, when far from the minimum~\cite{bertsekas:igd}. However, when close to the minimum, incremental SGD requires diminishing step sizes in order to converge. This translates into an additional hyper-parameter, i.e., the learning rate.

%%%%%%%%%%%%%%%%%%%%%%%%%%%%%%%%%%%%%
\begin{figure*}[h]
\begin{minipage}{.42\textwidth}
\underline{\textbf{Algorithm 2} Batch SGD Optimization Epoch}
\algsetup{linenodelimiter=.}

\begin{algorithmic}[1]

\STATE Compute gradient:\\
	\textbf{for} $i=1$ \textbf{to} $N$ \textbf{do}
$\vec{g} \leftarrow \vec{g} + {\overrightarrow{\nabla f} \left(\vec{w}; \vec{x_{i}}, y_{i} \right)}$
\STATE Update model: $\vec{w} \leftarrow \vec{w} - \alpha \vec{g}$

\end{algorithmic}
\end{minipage}
\hfill
\begin{minipage}{.52\textwidth}
\underline{\textbf{Algorithm 3} Incremental SGD Optimization Epoch}
\algsetup{linenodelimiter=.}

\begin{algorithmic}[1]

\STATE \textbf{for} $i=1$ \textbf{to} $N$ \textbf{do}
\STATE \hspace*{.25cm} Compute gradient estimate: $\vec{g} \leftarrow {\overrightarrow{\nabla f} \left(\vec{w}; \vec{x_{i}}, y_{i} \right)}$
\STATE \hspace*{.25cm} Update model: $\vec{w} \leftarrow \vec{w} - \alpha \vec{g}$
\STATE \textbf{end for}

\end{algorithmic}
\end{minipage}
\end{figure*}
%%%%%%%%%%%%%%%%%%%%%%%%%%%%%%%%%%%%%

%%%%%%%%%%%%%%%%%%%%%%%%%%%%%%%%%%%%%%%%%%%%%%%
\subsection{Synchronous Parallel SGD}\label{ssec:sgd:par-synch}

Parallelizing SGD seems rather impossible because of a chain dependency on model updates across epochs (batch) and even inside an epoch (incremental), where the current gradient relies on the previous model. As a result, in order to preserve the theoretical soundness of the algorithm, concurrency is limited to within each stage -- gradient computation and model update, respectively -- while synchronization has to be strictly enforced between them. Of the two stages, gradient computation entails significantly more work and is, thus, the main candidate for parallelization. This is achieved by expressing the gradient as a linear algebra formula over the training data and the model. As a concrete example, we give the linear algebra expression for the LR gradient:
\begin{equation}\label{eq:LR-grad:la}
\vec{g} = \vec{\vec{X}}_{k}^{T} \odot \left( -\vec{Y}_{k} \cdot \frac {e^{-\vec{Y}_{k} \cdot \vec{\vec{X}}_{k} \odot \vec{w}}} {1 + e^{-\vec{Y}_{k} \cdot \vec{\vec{X}}_{k} \odot \vec{w}}} \right)
\end{equation}
The algebraic operators involving vectors and matrices are element-wise when written in standard notation and have the linear algebra meaning, otherwise. For example, $\cdot$ corresponds to element-wise vector multiplication, while $\odot$ stands for vector dot-product or matrix-vector multiplication. This expression replaces the loop corresponding to gradient computation in \texttt{Batch SGD Optimization Epoch}. Efficient parallel implementations from highly-optimized linear algebra libraries are then used to speed-up the computation.

%%%%%%%%%%%%%%%%%%%%%%%%%%%%%%%%%%%%%%%%%%%%%%%
\subsection{Asynchronous Parallel SGD}\label{ssec:sgd:par-asynch}

In asynchronous parallel SGD, multiple model updates are executed concurrently---without synchronization primitives, e.g., mutexes or locks. Moreover, access to the model in gradient computation can also interfere with the update---or a stale model is used.  Hogwild~\cite{hogwild,bismarck,dimm-witted,hogbatch,cyclades,hogwild-disk} is the most representative algorithm in this category. It is the exact \texttt{Incremental SGD Optimization Epoch} with the loop executed in parallel: 
\begin{algorithmic}
\STATE \textbf{for} $i=1$ \textbf{to} $N$ \textbf{do \textit{in parallel}}
\end{algorithmic}
This makes the parallel implementation of Hogwild very simple---a single directive has to be added in OpenMP\footnote{\url{http://www.openmp.org/}}. Although not satisfying the specification of SGD, it has been theoretically proven that the resulting non-determinism in Hogwild enhances randomness and guarantees convergence for sparse models~\cite{hogwild}. However, due to parallelism, the overall time to convergence can be orders of magnitude faster than the sequential solution. The approach taken in synchronous SGD is more conservative---preserve the convergence of batch SGD exactly. ``Which of the two alternatives is better?'' is an open question that extends upon the debate between batch and incremental SGD---synchronous corresponds to batch, while asynchronous to incremental.

%%%%%%%%%%%%%%%%%%%%%%%%%%%%%%%%%%%%%%%%%%%%%%%%%%%%%%%%%%%%%%%%%
%\input{gpu-arch}

\section{COMPUTING ARCHITECTURES}\label{sec:architecture}

We present the two computing architectures considered in this work---multi-core NUMA CPU and GPU.

\begin{figure*}[htbp]
\begin{minipage}{.48\textwidth}
\vspace*{0cm}
\centering
\includegraphics[width=\textwidth]{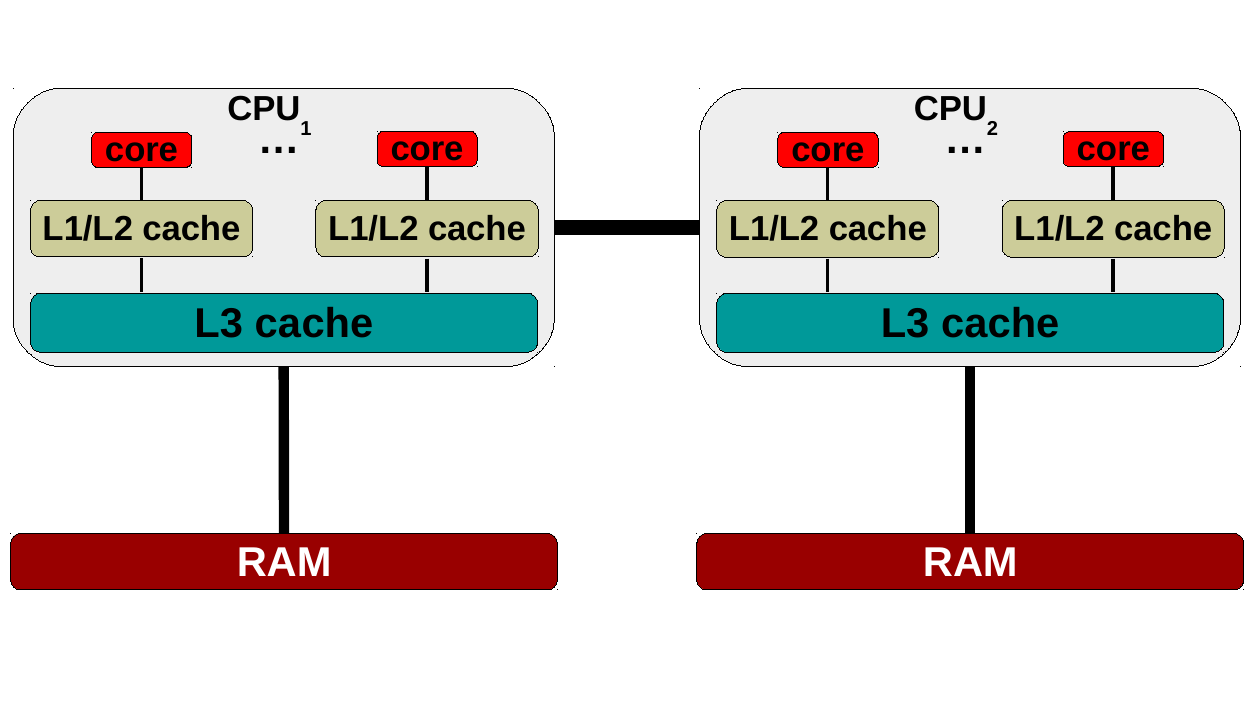}
\caption{NUMA CPU architecture.}
\label{fig:cpu-arch}
\end{minipage}
\begin{minipage}{.48\textwidth}
\centering
\includegraphics[width=\textwidth]{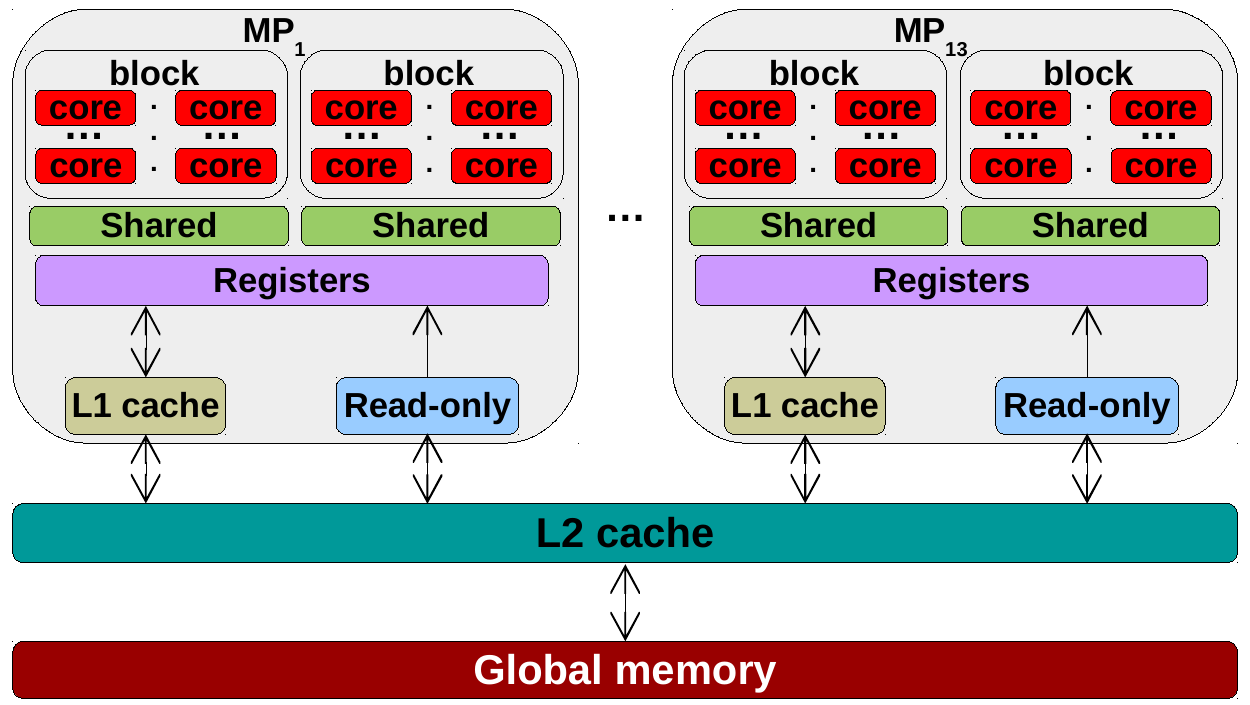}
\caption{GPU architecture.}
\label{fig:gpu-arch}
\end{minipage}
\end{figure*}

%%%%%%%%%%%%%%%%%%%%%%%%%%%%%%%%%%%%%%%
\textbf{NUMA CPU.}
The architecture of a NUMA machine is depicted in Figure~\ref{fig:cpu-arch}. It consists of several nodes which contain multiple cores and processor caches. The L1 and L2 caches are associated with each core, while the L3 cache is shared across all the cores in a node. Each node is directly connected to a region of the DRAM memory. NUMA nodes are connected to each other by high-bandwidth interconnects on the main board. To access DRAM regions of other nodes, data is transferred over these interconnects. However, this is slower than accessing the locally-associated memory. \textit{Cache-coherency is implicit} on NUMA machines and is implemented in hardware. In the worst case, the coherency protocol requires transfer across nodes which can generate congestion on the interconnect and, thus, significantly reduce the speedup of parallel solutions.

%%%%%%%%%%%%%%%%%%%%%%%%%%%%%%%%%%%%%
%%%%%%%%%%%%%%%%%%%%%%%%%%%%%%%%%%%%%

%%%%%%%%%%%%%%%%%%%%%%%%%%%%%%%%%%%%%%%
\textbf{GPU.}
As illustrated in Figure~\ref{fig:gpu-arch}, a GPU contains multiple streaming multiprocessors (MP). Each MP consists of a large number of specialized cores targeted at a limited subset of instructions. In the \textit{CUDA programming model}, work is issued to the GPU in the form of a function, referred to as the \textit{kernel}. A logical instance of the kernel is called a thread. The kernel code is parametrized by a logical thread identifier that allows each thread to operate on a different partition of the input data. Since thousands of threads can be executed concurrently across MPs, global thread synchronization is not available. Nonetheless, synchronization can be enforced at thread block level. A \textit{thread block (or block)} is a logical group of the threads executed for a kernel and imposes an upper limit on the number of threads, e.g., at most 1024 threads can be part of a block. The programmer has to specify both the number of blocks and the number of threads in a block when launching a kernel. Physically, all the threads in a block must reside on the same MP, as shown in Figure~\ref{fig:gpu-arch}. In order to highly utilize the GPU parallelism, the number of blocks has to be at least equal to the number of MP. To manage thousands of concurrent threads running on different parts of the data, the MP employs SIMT (single-instruction, multiple-thread) or SIMD (single-instruction, multiple-data) parallelism by grouping consecutive threads of a block into a \textit{warp}. The MP issues instructions at warp level in vector-like fashion for all the threads in the warp at a time. The number of threads in a block should be a multiple of the warp size in order to achieve full warp utilization. Moreover, in order to fully utilize all the cores, there should be a sufficiently large number of warps that contain sufficiently long sequences of independent instructions.

Threads can access the various units of the deep memory hierarchy in Figure~\ref{fig:gpu-arch} explicitly -- in the code -- during execution. This is quite different from the CPU memory management which is completely hidden from the programmer. While more flexible, it also makes GPU programming harder. Global memory (or device RAM memory) is persistent over multiple kernel invocations and can be accessed from all the threads across MPs. While the largest in size, global memory has the highest latency and lowest bandwidth. Shared memory is a low-latency high-bandwidth memory available to all the threads within a thread block. It is the only mechanism that allows synchronization between threads---only within a thread block, though. The read-only constant texture memory (or scratchpad memory) is a read-only cache populated from global memory. It is accessible by all the threads on an MP. Placement on the shared and scratchpad memory has to be implemented explicitly by the programmer. The two levels of cache L1 and L2 are used to improve the latency to the global memory. L1 cache handles only local thread memory and does not cache global memory loads. As a result, there is \textit{no cache coherency} implemented across MPs. When a global memory address is requested by a thread of a warp, aligned successive addresses are converted into a single memory transaction which is called \textit{memory coalescing}. To move data efficiently from global memory, the threads in a warp have to access consecutive global memory addresses. If the requested addresses of the warp are sparse or unaligned, several memory transactions are required to support the warp computations. Until all the requested data are cached in L2, the warp cannot be scheduled for computation.

%%%%%%%%%%%%%%%%%%%%%%%%%%%%%%%%%%%%%
\begin{table*}[htbp]
\begin{minipage}{.52\textwidth}
Table~\ref{tbl:hardware-spec} gives the hardware specifications of the NUMA machine and the NVIDIA Tesla K80 GPU used in this paper. While the number of cores and threads is much larger for the GPU, the numbers for the NUMA machine are quite high compared to previous CPU generations, e.g., 56 independent threads can run concurrently on a single machine. Although the amount of memory available on the CPU is 20X larger than on the GPU, the L2 cache on the GPU is 6X larger. This reflects the throughput emphasis of the GPU memory hierarchy as opposed to the latency optimization for CPU.
\end{minipage}
\hfill
\begin{minipage}{.4\textwidth}
  \begin{center}
    \begin{tabular}{l|rr}
    & \textbf{NUMA} & \textbf{GPU} \\
    \hline

		CPU/MP & 2 & 13 \\
		cores & 14 per CPU & 192 per MP \\
		blocks & - & 16 per MP \\
		threads & 28 per CPU & 2048 per MP \\
		L1 cache & 32+32 KB & 48 KB \\
		L2 cache & 256 KB & 1.5 MB \\
		L3/shared & 35 MB & 48 KB \\
		RAM/global & 256 GB & 12 GB \\
    \hline
    \end{tabular}
    \caption{Hardware specification.}\label{tbl:hardware-spec}
  \end{center}
\end{minipage}
\hfill
\end{table*}
%%%%%%%%%%%%%%%%%%%%%%%%%%%%%%%%%%%%%

%%%%%%%%%%%%%%%%%%%%%%%%%%%%%%%%%%%%%%%%%%%%%%%%%%%%%%%%%%%%%%%%%
%\input{synch-gd-gpu}

\section{SYNCHRONOUS SGD IMPLEMENTATION}\label{sec:synch-sgd}

The implementation of synchronous SGD consists of a sequence of primitive linear algebra function invocations for gradient computation and model update. In the case of the LR gradient (Eq.\eqref{eq:LR-grad:la}), the function sequence is:
\begin{sql}
vector a = matrix-vector-product(data, model)
a = vector-vector-element-product(label, a)
a = vector-element-exponent(a)
vector b = vector-element-sum(1, a)
a = vector-vector-element-division(a, b)
a = vector-vector-element-product(a, -label)
gradient = matrix-vector-product(transpose(data), a)
\end{sql}
Each of these functions is blocking, i.e., the next function in the sequence is invoked only after the previous finishes execution. This introduces a clear boundary between gradient computation and model update, essentially synchronizing access to the model. Parallelism is confined exclusively to intra-function processing. This allows for a variety of implementations, as long as the function API is preserved. ML frameworks, e.g., TensorFlow and BIDMach, capitalize on this abstraction and implement the linear algebra primitives with a unified API for both CPUs and GPUs. For example, function \texttt{matrix-vector-product} has an implementation for multi-thread CPU and one as a GPU kernel. Moreover, separate implementations are provided for dense and sparse data because of the different set of optimizations they require---this is not the case for TensorFlow and other deep learning frameworks which support only dense matrix representations and cannot, thus, handle high-dimensional models. The benefit of this approach is that switching between architectures does not require any code modification.

%%%%%%%%%%%%%%%%%%%%%%%%%%%%%%%%%%%%%
\begin{figure*}[htbp]
\centering
\includegraphics[width=\textwidth]{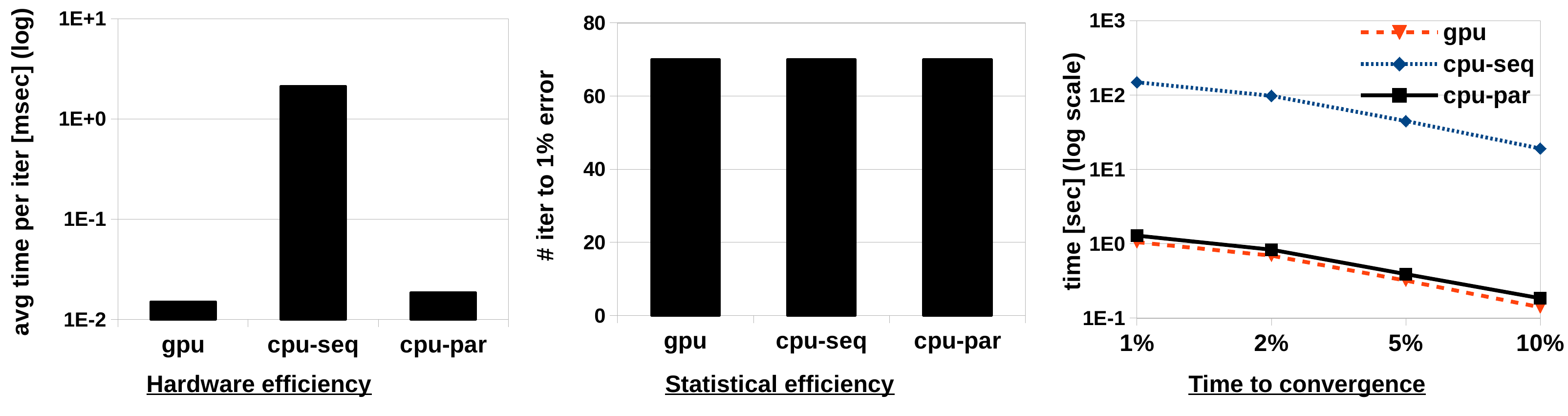}
\caption{Synchronous SGD on dense data (\texttt{covtype}).}
\label{fig:forest-synch}
\end{figure*}
%%%%%%%%%%%%%%%%%%%%%%%%%%%%%%%%%%%%%

Our synchronous SGD implementation follows the common API approach. We use the ViennaCL library which implements the linear algebra primitives used in LR and SVM---we extend the library with several functions. ViennaCL has support for multi-thread CPU and GPU, and for dense and sparse data. Since the ViennaCL implementations use all the specific architectural optimizations and are among the fastest available, we do not have to apply further intra-primitive optimizations. For a specific configuration, e.g., CPU or GPU, all the primitives are executed on that device---no cross-device execution. We do not apply cross-primitive optimizations such as pipelining, fusion, and on-GPU intermediate result caching~\cite{systemml} because this requires a holistic view of the program.

%%%%%%%%%%%%%%%%%%%%%%%%%%%%%%%%%%%%%
\begin{figure*}[htbp]
\centering
\includegraphics[width=\textwidth]{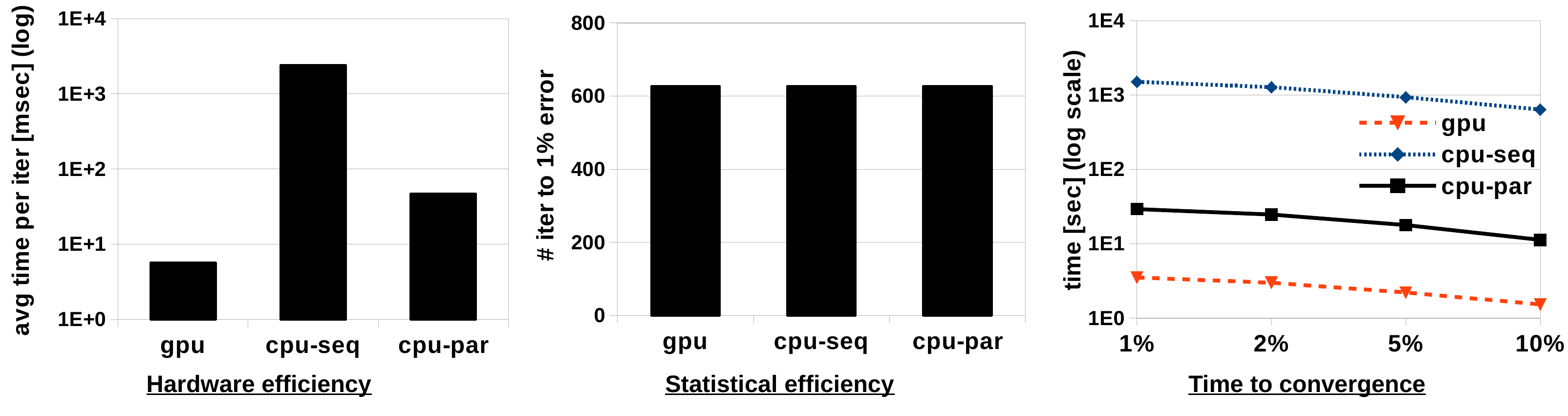}
\caption{Synchronous SGD on sparse data (\texttt{news}).}
\label{fig:news-synch}
\end{figure*}
%%%%%%%%%%%%%%%%%%%%%%%%%%%%%%%%%%%%%

%%%%%%%%%%%%%%%%%%%%%%%%%%%%%%%%%%%%%
The benefit of highly-optimized linear algebra primitives is reflected in the \textbf{hardware efficiency} of parallel synchronous SGD. Figure~\ref{fig:forest-synch} and~\ref{fig:news-synch} show that up to two orders of magnitude reduction is achieved over the sequential SGD for LR both by a multi-thread CPU and a GPU solution. These results also confirm that choosing the optimal kernel is highly-dependent on the data characteristics and the model. For dense data (Figure~\ref{fig:forest-synch}), the GPU kernel and parallel CPU are indistinguishable, while for sparse data (Figure~\ref{fig:news-synch}), the GPU is faster by an order of magnitude. The complete details of the datasets and the models are given in Section~\ref{sec:experiments}. Since the exact semantics of sequential SGD is preserved, the \textbf{statistical efficiency} is the same for multi-thread CPU and GPU (Figure~\ref{fig:forest-synch} and~\ref{fig:news-synch}). This property is important because the convergence analysis for sequential SGD can be immediately extended to the parallel synchronous SGD. Moreover, the \textbf{time to convergence} is determined entirely by the hardware efficiency. Figure~\ref{fig:forest-synch} and~\ref{fig:news-synch} confirm this linear dependency---the time to convergence is the number of epochs multiplied with the time of an epoch. As a result, GPU has faster convergence for sparse data, while on dense data, CPU and GPU are almost the same. Deciding the architecture on which to execute parallel synchronous SGD is not a straightforward decision, even though these results suggest that GPU is better. The convergence time is conditioned by the size and sparsity of the data, and the model. There is no simple rule on how to undoubtedly choose between CPU and GPU in ViennaCL---the same conclusion is supported by experiments with TensorFlow and BIDMach.

%%%%%%%%%%%%%%%%%%%%%%%%%%%%%%%%%%%%%%%%%%%%%%%%%%%%%%%%%%%%%%%%%
%\input{asynch-gd-gpu}

\section{ASYNCHRONOUS SGD IMPLEMENTATION}\label{sec:asynch-sgd}

Compared to the non-intrusive synchronous approach which composes a series of architecture-optimized linear algebra functions, the asynchronous solution consists of a single function that implements the \texttt{Incremental SGD Optimization Epoch}. For each training example, this function first computes the gradient and immediately applies it to a model update. Parallelism is achieved by executing several instances of the function, i.e., threads, concurrently over partitions of the examples. Asynchronous execution is the result of unprotected access to the shared state, i.e., the model, across concurrent threads. This approach to incremental SGD -- the Hogwild algorithm~\cite{hogwild} -- prioritizes hardware over statistical efficiency and better time to convergence is often obtained. However, a naive implementation is not sufficient~\cite{hogbatch}. Rather, architectural optimizations of the hardware have to be carefully considered.

%%%%%%%%%%%%%%%%%%%%%%%%%%%%%%%%%%%%%%%%%%%
\subsection{Hogwild on NUMA CPU}\label{ssec:async-sgd:cpu}

In DimmWitted~\cite{dimm-witted}, Zhang and Re give a Hogwild implementation optimized for NUMA CPU architectures. They investigate the impact of three factors -- access method, model replication, and data replication -- on the efficiency of Hogwild. For each factor, multiple alternatives are independently evaluated in terms of statistical and hardware efficiency and the optimal configuration is selected for every dataset/model combination---while the goal of DimmWitted is to do this automatically with a cost-based optimizer, unfortunately, it is not possible. Given the exhaustive nature of the DimmWitted study, our asynchronous SGD NUMA CPU implementation follows their solution. According to Figure 14 in~\cite{dimm-witted}, fully-replicated row-wise data access to a model replicated at NUMA node granularity is the optimal configuration for LR and SVM tasks. In our experimental setting, this corresponds to having two copies of the data and the model---one for each NUMA node. Essentially, we execute two independent Hogwild instances---one for each NUMA node. Due to the non-deterministic assignment of examples to threads and the model update order, these two instance are not identical copies---they provide non-redundant statistical information which improves the statistical efficiency. A second layer asynchronous Hogwild is executed between the NUMA node models in order to reduce their divergence. Periodically, a separate thread reads these two models, merges them, and updates each replica. Since all the threads scheduled on a NUMA node access only the corresponding model replica, no cache coherence overhead is incurred. This is reflected in higher hardware efficiency. For the remainder of the paper, all the references to asynchronous SGD on NUMA CPU correspond to this implementation.

%%%%%%%%%%%%%%%%%%%%%%%%%%%%%%%%%%%%%%%%%%%
\subsection{Hogwild on GPU}\label{ssec:async-sgd:gpu}

In this paper, we provide the first in-depth study of Hogwild on GPU. While several GPU extensions to asynchronous SGD have been proposed in the literature~\cite{SVM-CF-SGD-GPU,MF-SGD-GPU,CuMF-SGD}, they either target a single application, e.g., low-rank matrix factorization, or restrict themselves to a specific data/model configuration, e.g., sparse round-robin partitioned data with a single shared model. Moreover, none of these solutions is completely asynchronous in a Hogwild sense. We take an exhaustive approach in which we organize the Hogwild design space into three dimensions and consider several strategies for each dimension. Given a training dataset, a model specification, and a GPU configuration of threads and blocks, our goal is to determine an execution plan for the asynchronous Hogwild. The execution plan has to specify the assignment of data and model to GPU threads and the access path to the assigned data and model---inside the thread. Table~\ref{tbl:asynch-sgd:design-space} summarizes the components of the execution plan -- the dimensions of the design space -- and their corresponding strategies.
\begin{table*}[htbp]
\begin{minipage}{.4\textwidth}
Although derived from the NUMA CPU factors proposed in DimmWitted~\cite{dimm-witted}, the GPU optimizations are quite different because of the layered parallelism consisting of blocks and threads, the SIMD execution within a warp, and the distinct memory hierarchy optimized for throughput rather than latency. While our initial goal has been to build an analytical model that identifies the optimal execution plan for any data/model configuration automatically, we have found experimentally that this choice is highly-dependent on data and model characteristics. Thus, we limit ourselves to providing practical rules of thumb to guide the optimal choice.
\end{minipage}
\hfill
\begin{minipage}{.56\textwidth}
  \begin{center}
    \begin{tabular}{l|l}

	\textbf{Dimension} & \textbf{Strategies} \\
	
	\hline
	
	\multirow{4}{*}{Data access path} & row-major round-robin (row-rr) \\
	& row-major chunking (row-ch) \\
	& column-major round-robin (col-rr) \\
	& column-major chunking (col-ch) \\

	\hline

	\multirow{4}{*}{Model replication} & kernel \\
	& block \\
	& thread \\
	& example \\

	\hline

	\multirow{2}{*}{Data replication} & no replication (no-rep) \\
	& k-wise replication (rep-2, rep-5, rep-10)

    \end{tabular}
  \end{center}
\caption{Design space for Hogwild on GPU.}\label{tbl:asynch-sgd:design-space}
\end{minipage}
\end{table*}
%%%%%%%%%%%%%%%%%%%%%%%%%%%%%%%%%%

%%%%%%%%%%%%%%%%%%%%%%%%%%%%%%%%%%%%%%%%%%%%%%%%%%%%%
\subsubsection{Data Access Path}\label{ssec:asynch:data-access}

The training examples $\vec{x}_{i}$ form a 2-D matrix which can be organized in memory in row-major (row) -- by example -- or column-major (col) -- by feature -- format. These are extended to sparse data by the corresponding Compressed Sparse Row (CSR) and Compressed Sparse Column (CSC) format which store only the non-zero entries using three 1-D arrays. There are two arrays for the non-zero values and their column (or row) index -- each of size the number of non-zero entries -- and a third array with size the number of rows (or columns) that stores the index in the first two arrays where each column (row) starts. The assignment of examples to threads -- data partitioning -- is a second factor that has to be carefully considered when accessing the training data. The two standard approaches that do not require preprocessing and dynamic scheduling are round-robin (rr) and chunking (ch)---horizontal partitioning. The combination of data format and partitioning results in four configurations---row-rr, row-ch, col-rr, and col-ch. Existing NUMA CPU implementations consider only row-ch since asynchronous SGD accesses a complete example to compute the gradient and chunking allows for exclusive access to the local memory. SIMD execution coupled with memory access coalescing inside a thread block/warp pose new challenges for GPU. If the number of examples assigned to the threads of a block is not identical, the threads having fewer examples are stalled until the others finish. For dense data, all the threads in a block access the same model feature simultaneously which results in non-coalescing. This also happens for sparse data, however, the reason is the random access pattern to the model features. In both cases, the number of memory transactions necessary to execute a SIMD instruction is a bottleneck for the NUMA CPU row-ch configuration. Thus, investigating the other alternatives is essential.

\begin{figure}[htbp]
\begin{center}
\includegraphics[width=.98\textwidth]{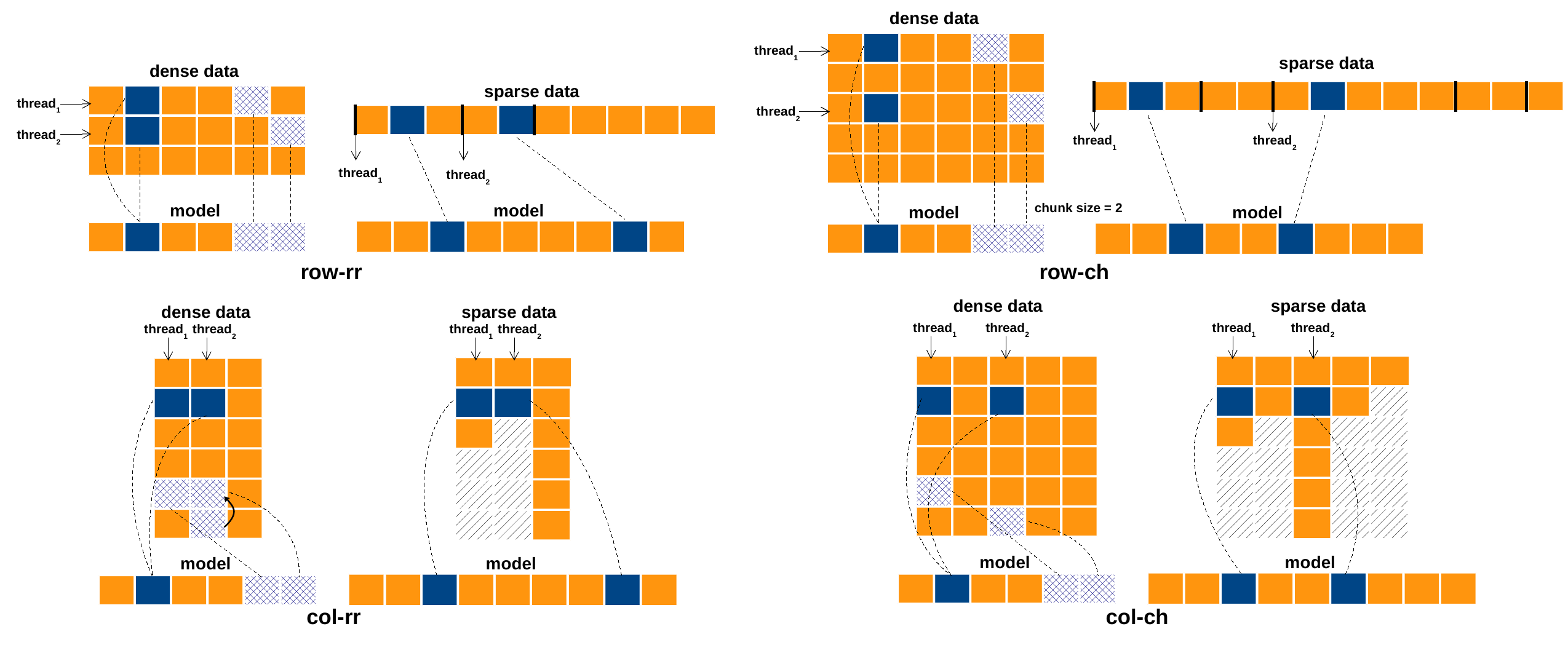}
\caption{Data access path configurations on GPU.}\label{fig:data-access-config}
\end{center}
\end{figure}

Figure~\ref{fig:data-access-config} depicts a graphical representation of the four data access path configurations considered for the execution on GPU: row-rr, row-ch, col-rr, and col-ch. The dashed lines link the features accessed from the dataset and updated in the model. In the dense dataset, the hatched cells indicate the optimization method we apply to reduce the data conflicts on the model updates. The hatched cells in the sparse dataset are the padded zero features used to convert the sparse dataset into a dense representation.

%%%%%%%%%%%%%%%%%%%%%%%%%%%%%%%%%%%%
%%%%%%%%%%%%%%%%%%%%%%%%%%%%%%%%%%%%%

%%%%%%%%%%%%%%%%%%%%%%%%%%%%%%%%%%%%%
\textit{row-rr.}
The training examples are stored in succession and they are accessed by consecutive threads in a block/warp. For dense data, the threads of a warp read the same feature from their example when computing the gradient and modify exactly the same feature of the shared model for every update. Therefore, only a single thread successfully updates the shared model which impacts statistical efficiency negatively. Moreover, the access pattern to examples and the model is not coalesced---consecutive threads do not access consecutive memory addresses. To reduce conflicts to the shared model, each thread starts the model update process from a different index, as shown by the hatched cells in Figure~\ref{fig:data-access-config}. The starting index is assigned circularly based on the thread id. As a result, when the model size is at least half of the warp size, no update conflicts exist. This optimization also provides coalesced access to the model, thus reducing the number of memory transactions. For sparse data, the number of features across examples varies and the feature indexes are discontinuous. When threads request the features at the same position, their index is likely different, leading to independent updates.

In addition to the model, every thread has a local gradient which is stored by default in the local memory. Since local memory is mapped to L1 cache, each feature occupies a 128-byte cache line. A thread has to fetch aligned data -- including nearby unrequested features -- which wastes most of the space in a cache line for loading a single feature. The warp also jumps to non-coalesced addresses multiple times to combine the separate requested features. With half-a-warp, i.e., 16 threads, every feature requests 16 cache lines which are 2048 bytes. We consider the alternative of storing the gradients in the global memory. In this case, a feature occupies a 32-byte segment in L2 cache; then 16 features request 16 cache segments which take only 512 bytes. While each feature still caches nearby unrequested indexes, the size of memory transactions to global memory is 4 times lower than to local memory. This reduction improves hardware efficiency.

%%%%%%%%%%%%%%%%%%%%%%%%%%%%%%%%%%%%%
\textit{row-ch.}
The training examples are stored consecutively and a thread also processes consecutive examples (Figure~\ref{fig:data-access-config}). Examples that are chunk size apart are processed concurrently by threads in a warp. Since these are minor differences compared to row-rr, the same optimizations apply. Nonetheless, depending on the chunk size and the number of threads, the performance can be quite different (Figure~\ref{fig:news-access}).

%%%%%%%%%%%%%%%%%%%%%%%%%%%%%%%%%%%%%
\textit{col-rr.}
To implement the column-major representation, we transpose the training examples. This coalesces features across examples in global memory (Figure~\ref{fig:data-access-config}) and the data accessed by a warp covers full segments. On average, each feature takes only a quarter of a memory transaction to be cached. In order to avoid update conflicts, threads access the model in circular order---similar to row-rr. The CSR sparse format is, however, not adequate for column-major access because we have to traverse the 1-D array which records the row indices of non-zero features to locate all the features of an example. Therefore, we map sparse data into a dense padded format that stores all the examples at the same width---equal to the maximum number of non-zero features. When the threads of a warp request features from the same position, accesses are coalesced. Moreover, the shared model is updated without conflict.

%%%%%%%%%%%%%%%%%%%%%%%%%%%%%%%%%%%%%
\textit{col-ch.}
As in the case of row-major, the main difference between col-ch and col-rr is the width between examples processed concurrently in a warp. This impacts the number of memory transactions directly. When the chunk size is small -- due to a large number of threads -- the average memory transaction can be much smaller than 1 since features share a cache segment. For sparse data, the number of features is the other crucial factor impacting memory transactions. If the number of features differs greatly across the threads in a warp, the average number of memory transactions is closer to the maximum of 1.

%%%%%%%%%%%%%%%%%%%%%%%%%%%%%%%%%%%%%
\begin{figure*}[htbp]
\centering
\includegraphics[width=\textwidth]{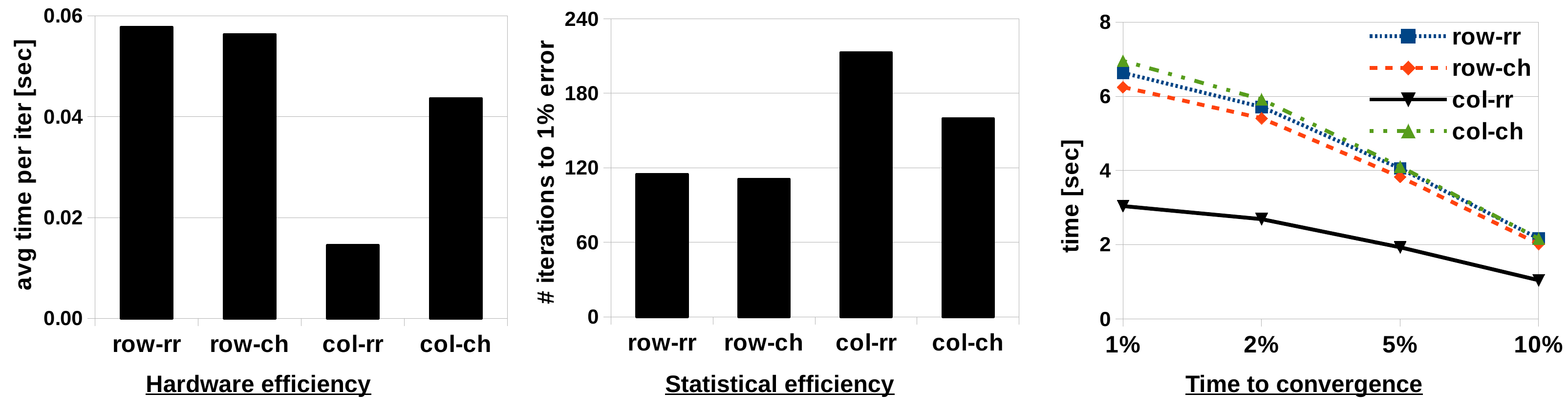}
\caption{Access path selection for dense data (\texttt{covtype}).}
\label{fig:forest-access}
\end{figure*}
%%%%%%%%%%%%%%%%%%%%%%%%%%%%%%%%%%%%%

%%%%%%%%%%%%%%%%%%%%%%%%%%%%%%%%%%%%%
\textbf{Hardware efficiency.}
Figure~\ref{fig:forest-access} and~\ref{fig:news-access} depict the performance of the data access strategies for LR. For dense data, col-rr exhibits the highest hardware efficiency, taking minimum average time per iteration. Since the gradient is stored in the global memory, the number of memory transactions is mostly determined by the number of L2 cache segments accessed. col-rr takes full advantage of cache access coalescing. For sparse data,  row-rr is slightly better than col-rr due to extra access to padded zero features. Moreover, the non-zero features at the same position in the CSR format are more likely to be in the same cache segment. Although round-robin access tends to perform better than chunking, this cannot be generalized since row-ch has better hardware efficiency than row-rr on dense data.

%%%%%%%%%%%%%%%%%%%%%%%%%%%%%%%%%%%%%
\textbf{Statistical efficiency.}
The strategy with the highest hardware efficiency has the lowest statistical efficiency in Figure~\ref{fig:forest-access} and~\ref{fig:news-access}. We find that the behavior of cache coalescing has a negative effect on model update conflicts. When data requests are coalesced, the warps are not frequently stalled for memory transactions. We can have at most 52 concurrent warps -- 1664 threads in total -- scheduled by the warp schedulers for computation. Such a large number of threads updating the model concurrently increases the conflicts to the single shared model. As a result, fewer updates survive which results in reduced statistical efficiency and more iterations to converge.

%%%%%%%%%%%%%%%%%%%%%%%%%%%%%%%%%%%%%
\begin{figure*}[htbp]
\centering
\includegraphics[width=\textwidth]{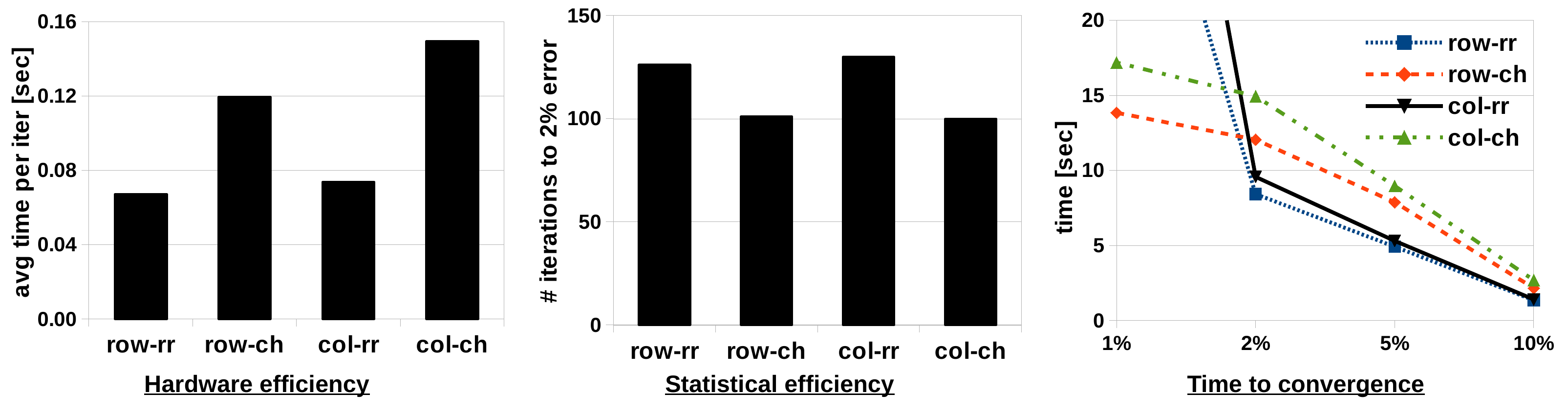}
\caption{Access path selection for sparse data (\texttt{news}).}
\label{fig:news-access}
\end{figure*}
%%%%%%%%%%%%%%%%%%%%%%%%%%%%%%%%%%%%%

%%%%%%%%%%%%%%%%%%%%%%%%%%%%%%%%%%%%%
\textbf{Time to convergence.}
Unlike synchronous SGD, both hardware and statistical efficiency have to be considered for analyzing time to convergence. As shown in the figures, these are inversely correlated which makes reasoning more difficult. On dense data, the reduction in time per iteration obtained by col-rr is sufficient to outperform the larger number of iterations and result in faster time to convergence. On sparse data, however, the excessive number of model update conflicts introduced by higher hardware efficiency prohibits optimal convergence. The round-robin strategies -- although converging faster to 2\% error -- cannot converge to 1\% error. Overall, the \textit{rule of thumb is col-rr data access with circular model updates is the optimal strategy}---with the caveat that optimal convergence on sparse data is harder to achieve.

%%%%%%%%%%%%%%%%%%%%%%%%%%%%%%%%%%%%%%%%%%%%%%%%%%%%%
\subsubsection{Model Replication}\label{ssec:asynch:model-replica}

As the shared model is accessed by multiple threads simultaneously, this can lead to degradation in statistical efficiency. In the case of GPU, this problem is triggered by the SIMD execution inside a warp -- the threads in a warp read/write the model exactly at the same time -- not by a cache coherency mechanism. Among all the threads in a warp which update the model simultaneously, only a single value is arbitrarily picked for each feature, while the other updates are discarded. Moreover, a warp may execute with a stale model -- due to stalling and rescheduling caused by cache misses -- since other warps keep modifying the model in the shared L2 cache. Given the inefficiencies incurred by a shared model per \textit{kernel}, we consider several alternatives that reduce sharing by exploiting the deep GPU memory hierarchy.

%%%%%%%%%%%%%%%%%%%%%%%%%%%%%%%%%%%%
\begin{figure}[htbp]
\begin{center}
\includegraphics[width=.88\textwidth]{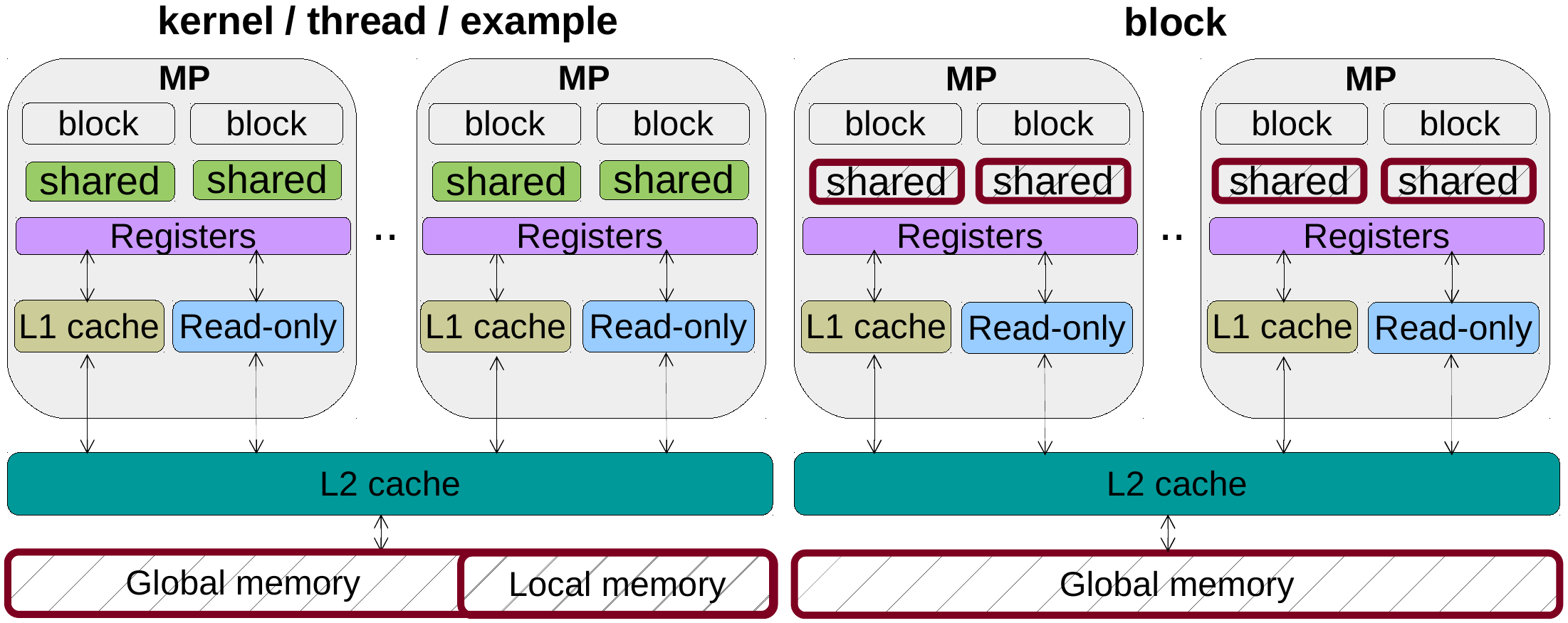}
\caption{Model replication configurations on GPU.}\label{fig:model-replica-config}
\end{center}
\end{figure}
%%%%%%%%%%%%%%%%%%%%%%%%%%%%%%%%%%%%%

Figure~\ref{fig:model-replica-config} shows the memory layout of the model replication configurations on GPU: kernel, block, thread, and example. In our design, the examples of the dataset are transferred into the GPU global memory while the model replicas are stored in different levels of the memory. We allocate space in DRAM for kernel, thread, and example and compare the difference between the performance of the allocations from global and local memory. As we analyze for gradient storage in Section~\ref{ssec:asynch:data-access}, although the data in local memory are cached to on-chip L1 with 128-byte lines, most of the non-coalesced data are not requested for computation. The kernel strategy has a single model shared by all the threads on the multiprocessors. In the thread strategy, the sizes of the model replicas are not identical since we materialize only the features accessed by every thread. While the example strategy maintains the model replicas with the same features for every example, the size of the model replicas are also different. The model replicas of the block strategy are stored in shared memory. If the model size is larger than the size of shared memory per thread-block, the model replicas are spilled to global memory.

%%%%%%%%%%%%%%%%%%%%%%%%%%%%%%%%%%%%%
\textit{block.}
There is a model replica for each block which is initialized before each iteration and merged into the global model at the end of the iteration---CUDA supports block-level synchronization. If the model is small enough, it is allocated in the shared memory---sliced into a number of 64-bit wide banks. Bank conflicts are avoided since each feature is accessed independently by a thread. Otherwise, threads incurring a bank conflict have to execute serially. Moreover, if a warp accesses the same feature while updating and copying the model replica, conflicts can be avoided altogether because of broadcasting of features in the same bank. Large models are replicated in the global memory---one per block. Initializing and merging model replicas incur additional overhead.

%%%%%%%%%%%%%%%%%%%%%%%%%%%%%%%%%%%%%
\textit{thread.}
There is a model replica which resides in local memory for each thread. For small models, this is mapped into the registers and provides fast access. Otherwise, it is mapped into the global memory. There are no model update conflicts since each thread has its own replica. Nonetheless, merging the replicas incurs significant overhead due to the large number of models.

%%%%%%%%%%%%%%%%%%%%%%%%%%%%%%%%%%%%%
\textit{example.}
To avoid merging sparse models in CSR format, we introduce example replication in which every example has its own model replica in global memory. An example replica contains the same features as the example. After a thread finishes processing the assigned examples, it copies the model replicas to the shared model. example replication avoids conflicts between threads almost entirely at the expense of memory usage.

%%%%%%%%%%%%%%%%%%%%%%%%%%%%%%%%%%%%%
\begin{figure*}[htbp]
\centering
\includegraphics[width=\textwidth]{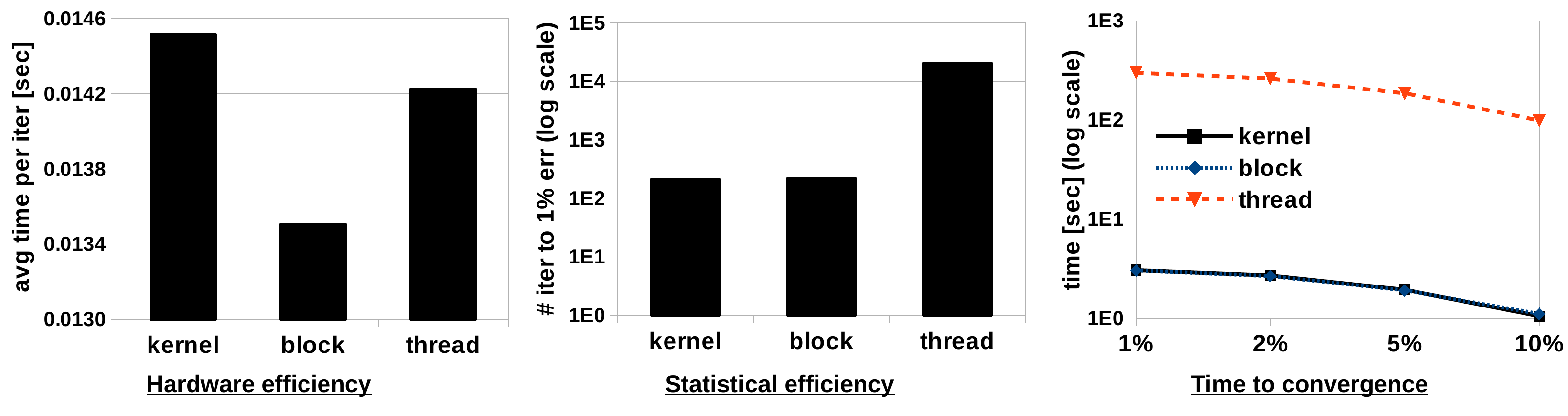}
\caption{Model replication effect on dense data (\texttt{covtype}).}
\label{fig:forest-model}
\end{figure*}
%%%%%%%%%%%%%%%%%%%%%%%%%%%%%%%%%%%%%

\begin{figure*}[htbp]
\centering
\includegraphics[width=\textwidth]{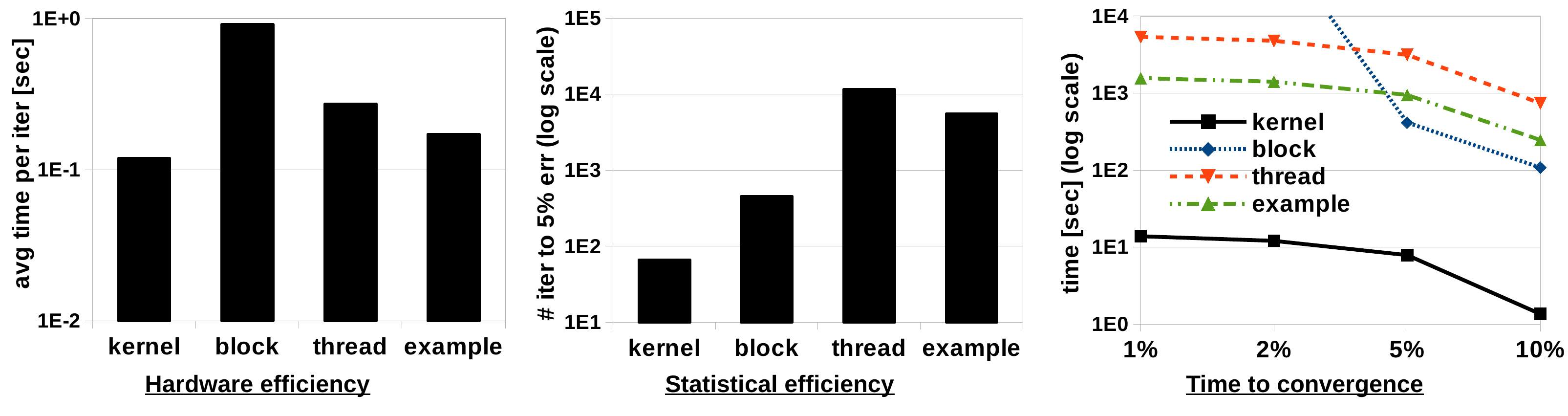}
\caption{Model replication effect on sparse data (\texttt{news}).}
\label{fig:news-model}
\end{figure*}

%%%%%%%%%%%%%%%%%%%%%%%%%%%%%%%%%%%%%
\textbf{Hardware efficiency.}
Figure~\ref{fig:forest-model} and~\ref{fig:news-model} plot the results with the col-rr access path for dense data---example replication applies only to sparse data. On dense data, block replication has the highest hardware efficiency because access to shared memory is faster than to global memory. thread has higher hardware efficiency than kernel, even though it incurs overhead to merge the local model. The reason is that the local models are cached in L1 which resides on MP, while in kernel the model is cached in L2 which resides off-chip. On sparse data, all the strategies utilize global memory to maintain the local and global models. kernel replication has the highest hardware efficiency, while block has the lowest---by a factor of 10. The synchronization required in block replication delays the threads with fewer non-zero features until the largest example is processed. example replication has no synchronization and caches models in L1, rather than L2---the case for thread and kernel.

%%%%%%%%%%%%%%%%%%%%%%%%%%%%%%%%%%%%%
\textbf{Statistical efficiency.}
kernel replication displays the highest statistical efficiency both for dense and sparse data. thread replication is two orders of magnitude lower. Intuitively, the more replicas, the lower the statistical efficiency. Model replicas -- whether stored on-chip or off-chip -- record only partial information. When update conflicts occur, features in the model replicas are discarded which leads to the shared model fetching limited information. In addition, the dot-product computed with the local replica generates an unreliable gradient---especially for sparse data.

%%%%%%%%%%%%%%%%%%%%%%%%%%%%%%%%%%%%%
%%%%%%%%%%%%%%%%%%%%%%%%%%%%%%%%%%%%%

%%%%%%%%%%%%%%%%%%%%%%%%%%%%%%%%%%%%%
\textbf{Time to convergence.}
On dense data, kernel and block have similar time to convergence, while thread is completely outperformed because of its low statistical efficiency. On sparse data, kernel converges the fastest---by two orders of magnitude. block fails to converge to 1\% error in an acceptable amount of time. Although in this case example is second, for different access path configurations, example outperforms kernel in time to convergence. Overall, \textit{the rule of thumb is to choose kernel model replication}. However, if the model fits in the shared memory, block replication may be a better choice.

%%%%%%%%%%%%%%%%%%%%%%%%%%%%%%%%%%%%%%%%%%%%%%%%%%%%%
\subsubsection{Data Replication}\label{ssec:asynch:data-replica}

Standard data partitioning does not replicate examples across threads because generating the correct query result requires adequate post-processing which increases execution time. We adopt this \textit{no replication (no-rep)} strategy in the data access path methods. Model training, however, is approximate and using more data has the potential to improve statistical efficiency. In order to explore this tradeoff between hardware and statistical efficiency, we investigate \textit{k-wise replication} in which the example partitions assigned to threads overlap at the boundaries. For example, \textit{2-wise replication (rep-2)} assigns two additional examples to every thread. The reason behind replication at the boundary is to preserve coalesced memory access. k-wise replication interacts minimally with model replication. The only difference is that the number of examples assigned to a thread is larger by k. 

%%%%%%%%%%%%%%%%%%%%%%%%%%%%%%%%%%%%
\begin{figure}[htbp]
\begin{center}
\includegraphics[width=\textwidth]{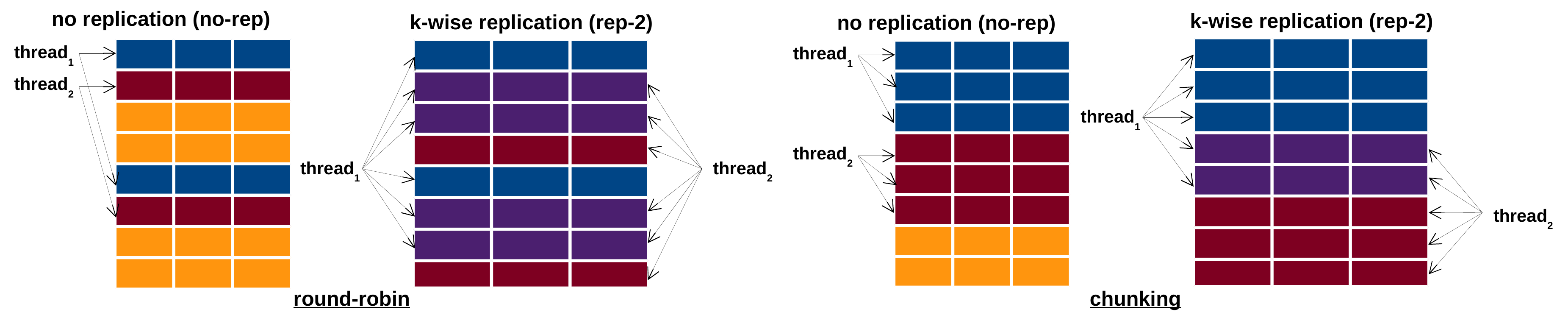}
\caption{Data replication configurations on GPU.}\label{fig:data-replica-config}
\end{center}
\end{figure}
%%%%%%%%%%%%%%%%%%%%%%%%%%%%%%%%%%%%%

Figure~\ref{fig:data-replica-config} plots the data replication configurations with two horizontal partitioning strategies: no replication (no-rep) and k-wise replication (rep-2). Without replication, the examples are sliced by the round-robin access path where threads take turn to process the examples one by one. With the chunking access path, the fixed-size chunks are assigned to the threads sequentially. In the k-wise replication strategy, the thread fetches several examples after the originally-assigned ones with no-rep, as Figure~\ref{fig:data-replica-config} shows. Therefore, the examples are accessed several times by different threads.

%%%%%%%%%%%%%%%%%%%%%%%%%%%%%%%%%%%%%
\begin{figure*}[htbp]
\centering
\includegraphics[width=\textwidth]{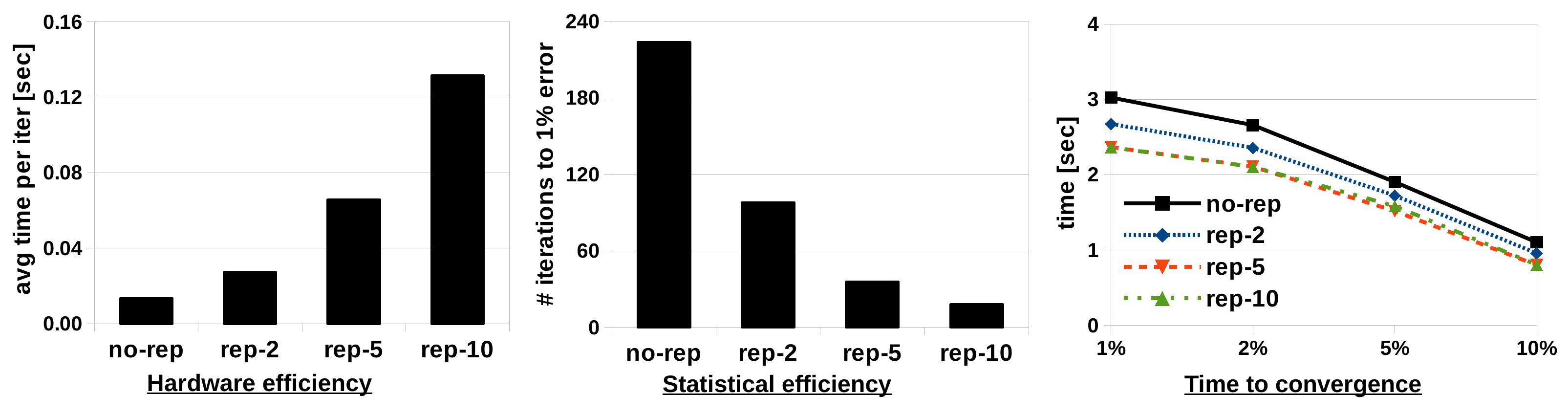}
\caption{Data replication effect on dense data (\texttt{covtype}).}
\label{fig:forest-data}
\end{figure*}
%%%%%%%%%%%%%%%%%%%%%%%%%%%%%%%%%%%%%

%%%%%%%%%%%%%%%%%%%%%%%%%%%%%%%%%%%%%
\begin{figure*}[htbp]
\centering
\includegraphics[width=\textwidth]{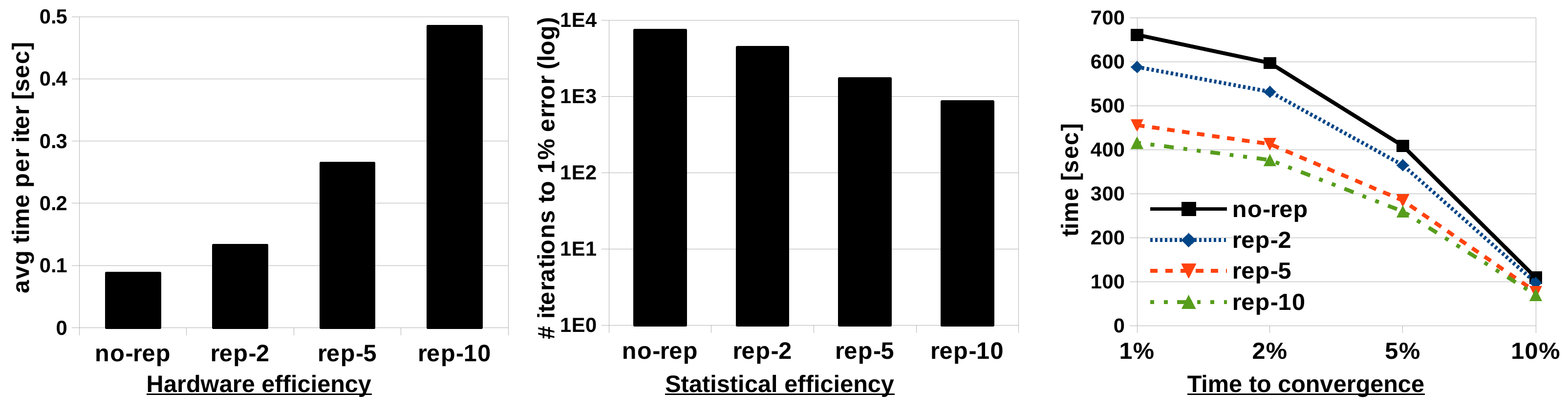}
\caption{Data replication effect on sparse data (\texttt{news}).}
\label{fig:news-data}
\end{figure*}
%%%%%%%%%%%%%%%%%%%%%%%%%%%%%%%%%%%%%

Figure~\ref{fig:forest-data} and~\ref{fig:news-data} depict the impact of k-wise replication on hardware and statistical efficiency for k equal to 2, 5, and 10, respectively. The trend is identical both on dense and sparse data. The larger k is, \textbf{hardware efficiency} drops almost linearly. This is somewhat unexpected because we expect coalesced memory access to be more resilient when the range increases only slightly. The configurations used in the figures are \textit{col-rr+kernel} (dense) and \textit{col-rr+example} (sparse). We observe similar trends for the other combinations. \textbf{Statistical efficiency} increases linearly with k since more information is extracted from the data for gradient computation and model update. When these opposing measures get combined in \textbf{time to convergence}, the reduction in number of iterations dominates the increase in time per iteration and we obtain faster convergence when the degree of replication is higher. The improvement diminishes beyond \textit{rep-10}. Overall, \textit{the rule of thumb is to adopt a limited degree of data replication} because it enhances statistical convergence linearly while also decreasing time to convergence by a significant margin.

%%%%%%%%%%%%%%%%%%%%%%%%%%%%%%%%%%%%%%%%%%%%%%%%%%%%%%%%%%%%%%%%%
%\input{experiments}

\section{EXPERIMENTAL EVALUATION}\label{sec:experiments}

We perform an extended empirical study across the exploratory axes defined in Figure~\ref{fig:study-axes} with respect to the performance axes introduced in Figure~\ref{fig:perf-axes}. The goal is to fully characterize the relationship between the considered configurations and understand the relevance of each performance measure. We validate our results by comparing against two representative analytics frameworks that have support both for CPU and GPU---TensorFlow and BIDMach. We emphasize that the main objective of the comparison is to add other reference points on the performance axes beyond our implementation, while the direct comparison between frameworks is secondary. Specifically, our experiments target the following questions:
\begin{compactitem}
\item What is role of the computing architecture, i.e., CPU/GPU, on the performance of synchronous SGD?
\item What is role of the computing architecture, i.e., CPU/GPU, on the performance of asynchronous SGD?
\item What is the optimal configuration for asynchronous SGD on GPU?
\item How do synchronous and asynchronous SGD compare against each other on CPU and GPU separately, and across computing platforms?
\item Are our implementations efficient with respect to TensroFlow and BIDMach?
\item How do the proposed algorithms scale with the number of training examples and the dimensionality of the feature vector, respectively?
\end{compactitem}

%%%%%%%%%%%%%%%%%%%%%%%%%%%%%%%%%%
\subsection{Setup}\label{sec:experiments:setup}

%%%%%%%%%%%%%%%%%%%%%%%%%%%%%%%%%%
\textbf{Implementation.}
We implement all the 8 configurations in Figure~\ref{fig:study-axes} following the best practices for Intel multi-core CPUs and for NVIDIA GPUs, respectively. We use OpenMP for multi-thread programming on the CPU and CUDA 8.0 on the GPU. Synchronous SGD is implemented using the ViennaCL (1.7.1) linear algebra library which provides optimized primitives with the same API for CPU and GPU. This allows us to have identical implementations---only compiled with different flags. We have separate implementations for dense and sparse data that use optimized data structures. All the code is written in \texttt{C++}. For the implementations in TensorFlow (0.12.0) and BIDMach (2.0.1), we write only the driver programs which define the objective function corresponding to the analytic model. We then invoke the synchronous SGD optimizer which calls the linear algebra kernels necessary in the gradient computation. While the driver is written in \texttt{python} for TensorFlow and \texttt{scala} for BIDMach, the linear algebra kernels are coded in \texttt{C++/CUDA} and are highly-optimized.

%%%%%%%%%%%%%%%%%%%%%%%%%%%%%%%%%%
\begin{table}[htbp]
  \begin{center}
    \begin{tabular}{l|rrrrr}

	\textbf{dataset} & \textbf{\#examples} & \textbf{\#features} & \textbf{\#nnz/example (avg)} & \textbf{sparse size} & \textbf{dense size} \\
	
	\hline

	covtype & 581,012 & 54 & 54 (54) & 485 MB & 485 MB\\
	w8a & 64,700 & 300 & 1 to 114 (11.65) & 4.4 MB & 155 MB \\
	real-sim & 72,309 & 20,958 & 1 to 3,484 (51.30) & 87 MB & 12.1 GB \\
	rcv1 & 677,399 & 47,236 & 4 to 1,224 (73.16) & 1.2 GB & 256 GB \\
	news & 19,996 & 1,355,191 & 1 to 16,423 (454.99) & 134 MB & 217 GB
    \end{tabular}
  \end{center}
\caption{Experimental datasets. nnz is number of non-zero.}\label{tbl:datasets}
\end{table}
%%%%%%%%%%%%%%%%%%%%%%%%%%%%%%%%%%

%%%%%%%%%%%%%%%%%%%%%%%%%%%%%%%%%%
\textbf{System.}
The properties of the computing architectures used in the experiments are presented in Figure~\ref{tbl:hardware-spec}. They are mounted in the same physical machine running Ubuntu 16.04 SMP with Linux kernel 4.4.0-77 and CUDA 8.0. Out of the two cards inside the Tesla K80 GPU, only one is used in the experiments. The two cards are seen as two independent GPUs by the operating system and have to be programmed independently. This is the default setting in both TensorFlow and BIDMach which require the programmer to specify the GPU on which the SGD optimizer executes. We plan to perform an evaluation with multiple -- possibly distributed -- GPUs in the future.

%%%%%%%%%%%%%%%%%%%%%%%%%%%%%%%%%%
\textbf{Methodology.}
We perform all the experiments at least 3 times and report the average value as the result. Each task is run for at least 10 iterations and the hardware efficiency is measured as the average execution time over the total number of iterations. The time to evaluate the loss is not included in the iteration time. All configurations/systems are initialized with the same model which gives the same initial loss. The SGD step size is chosen by griding its range in powers of 10, e.g., $\{10^{-6},10^{-5},\dots,10^{2}\}$, and selecting the value that generates the fastest time to convergence. The optimal step size varies across configurations. For end-to-end performance, we measure the wall-clock time it takes for each configuration to converge to a loss that is within 10\%, 5\%, 2\%, and 1\% of the optimal loss. Following prior work~\cite{dimm-witted}, we obtain the optimal loss by running all configurations for one hour and choosing the lowest. The time to load the data and output the result is not included. Moreover, in the case of GPU, the time to transfer the data and the model to/from the GPU global memory is also not included---we measure only the kernel execution time. For Hogwild on GPU, we report only the result corresponding to the configuration that achieves 1\% convergence error the earliest. As discussed in the results, this configuration is different across datasets.

%%%%%%%%%%%%%%%%%%%%%%%%%%%%%%%%%%
\textbf{Datasets and tasks.}
Rather than considering only two datasets -- dense and sparse -- we include five real datasets (Table~\ref{tbl:datasets}) that exhibit large variety in size, dimensionality, and sparsity. The number of dimensions varies from tens to more than 1 million, while the number of non-zero entries per example is as small as 1 for most of the datasets. In terms of physical size, the sparse representation is as small as 4.4 MB for \texttt{w8a} -- it can be cached (almost) completely both on CPU and GPU -- and as large as 1.2 GB for \texttt{rcv1}. In dense format, only \texttt{covtype} and \texttt{w8a} fit on the GPU, while \texttt{rcv1} and \texttt{news} cannot be processed even on the CPU. \texttt{covtype} is the representative dense dataset throughout the paper since it is complete, while \texttt{news} has the highest sparsity ratio of $3\times 10^{-4}$. These datasets have been used previously to evaluate the performance of parallel SGD on NUMA CPU~\cite{dimm-witted} and GPU~\cite{hogbatch}---more details can be found therein. We use a dense format for \texttt{w8a} in order to allow TensorFlow to execute. As a result, synchronous SGD becomes batch gradient descent since it uses the complete dataset to compute the exact gradient. Even if we reduce the batch size and have several model updates per data pass, the time to convergence does not improve. With five datasets and four points on the exploratory axes, we obtain 20 configurations per model. Since we consider two tasks -- LR and SVM -- we have 40 configurations overall. We do not include any regularization in the objective function in order to measure only the time spent in the actual computation.

%%%%%%%%%%%%%%%%%%%%%%%%%%%%%%%%%%
\subsection{Results}\label{sec:experiments:results}

We project the results on a subset of dimensions in the exploratory axes to facilitate a direct comparison between configurations. For synchronous and asynchronous updates taken separately, we compare the CPU and GPU implementations. Then we perform a direct comparison between the best synchronous and asynchronous configurations. For each computing architecture, we compare synchronous and asynchronous SGD, and the synchronous solutions in TensorFlow and BIDMach, respectively. Lastly, we study the effect of the number of examples and features on hardware efficiency for CPU and GPU separately.

%%%%%%%%%%%%%%%%%%%%%%%%%%%%%%%%%%
\begin{table*}[htb]
  \begin{center}
    \begin{tabular}{l|l||rrr|rrr|r}

	\multirow{2}{*}{\textbf{task}} & \multirow{2}{*}{\textbf{dataset}} & \multicolumn{3}{c|}{\textbf{time to convergence (sec)}} & \multicolumn{3}{c|}{\textbf{time per iteration (msec)}} & \multirow{2}{*}{\textbf{\# iterations}} \\

	& & \textbf{gpu} & \textbf{cpu-seq} & \textbf{cpu-par} & \textbf{gpu} & \textbf{cpu-seq} & \textbf{cpu-par} & \\

	\hline
	
	\multirow{5}{*}{LR} & covtype & \underline{1.05} & 145.11 & 1.29 & \underline{15} & 2,073 & 18.42 & 70 \\
	
	& w8a & \underline{0.37} & 148.88 & 0.46 & \underline{4.87} & 1,959 & 6.05 & 76 \\
	
	& real-sim & \underline{3.10} & 1,537.90 & 7.67 & \underline{4.43} & 2,197 & 10.96 & 700 \\
	
	& rcv1 & \underline{31.69} & 2,227.05 & 48.06 & \underline{44.82} & 3,150 & 67.98 & 707 \\

	& news & \underline{0.65} & 240.21 & 3.68 & \underline{6.37} & 2,355 & 36.08 & 102 \\

	\hline
	
	\multirow{5}{*}{SVM} & covtype & \underline{10.22} & 1,344.65 & 13.50 & \underline{14.27} & 1,878 & 18.85 & 716 \\
	
	& w8a & \underline{0.78} & 342.85 & 0.80 & \underline{4.13} & 1,814 & 4.23 & 189 \\
	
	& real-sim & \underline{0.23} & 75.59 & 0.46 & \underline{6.22} & 2,043 & 12.43 & 37 \\
	
	& rcv1 & \underline{1.13} & 111.61 & 2.61 & \underline{29.74} & 2,937 & 68.69 & 38 \\

	& news & \underline{0.30} & 98.42 & 1.69 & \underline{6.67} & 2,187 & 37.56 & 45 \\

    \end{tabular}
  \end{center}
\caption{Synchronous SGD performance to 1\% convergence error. The best values for each dataset are underlined.}\label{tbl:synch-bgd}
\end{table*}
%%%%%%%%%%%%%%%%%%%%%%%%%%%%%%%%%%

%%%%%%%%%%%%%%%%%%%%%%%%%%%%%%%%%%
\subsubsection{Synchronous SGD}\label{sec:experiments:results:synch-sgd}

Table~\ref{tbl:synch-bgd} contains the time to convergence, and hardware and statistical efficiency results for synchronous SGD implemented with the ViennaCL library. In order to show the improvement over a sequential solution, we also include results for a single-thread CPU implementation which achieves convergence in less than 5 minutes only for 6 out of the 10 dataset/task pairs. Since the parallel implementations always achieve convergence in less than 1 minute -- sometimes even in less than a second -- it is clear that parallelism helps. When comparing the multi-core CPU and GPU solutions, there is a clear trend---\textit{GPU is always faster than parallel CPU in time to convergence}. Since the statistical efficiency is identical in synchronous SGD independent of the computing platform, this also translates into faster GPU time per iteration, i.e., better hardware efficiency. Given that ViennaCL defines its internal representation for dense and sparse data and implements optimized kernels for CPU and GPU independently, the difference is due exclusively to the computational power of the two architectures---the GPU has more FLOPS than the CPU. The gap between the two architectures -- reflected by the speedup (Table~\ref{tbl:synch-sgd-speedup}) -- increases with the sparsity of the data. The GPU is faster by a small margin -- 32\% or less -- on the dense datasets \texttt{covtype} and \texttt{w8a}. This confirms that having powerful CPUs with tens of cores provides extensive parallelism that has the potential to close the gap on the GPU. In fact, we find that computing the transpose of a dense matrix on the GPU is a serious bottleneck in ViennaCL\footnote{\url{https://devblogs.nvidia.com/efficient-matrix-transpose-cuda-cc/}}. When the transpose is computed in-place, the GPU performance becomes worse than the CPU---in these results, the transpose is materialized and passed as a second matrix to the kernels. The gap between GPU and CPU increases on sparse data to more than 5X on \texttt{news}. Parallelizing linear algebra operations on sparse data is known to be a difficult task because of the irregular memory access~\cite{spmv-multicore}. This turns out to be more acute for the CPU memory hierarchy. At a close inspection, we find that ViennaCL takes advantage of the programmability of the GPU memory and exploits it to optimize coalesced access to sparse data. Moreover, the linear algebra kernels invoked depend on data sparsity---a different kernel is called on \texttt{news} compared to \texttt{real-sim} and \texttt{rcv1}. The high variance in statistical efficiency across dataset/task combinations is due to different hyper-parameters. It confirms that convergence rate is a property of both the task and the dataset and is independent of sparsity. When comparing the time per iteration across LR and SVM, however, we observe very similar results on GPU and CPU although their gradients are quite different. The reason for this is matrix batch processing and vectorization which hide the higher latency  incurred by individual element-wise operations---both LR and SVM have exactly the same number of matrix-matrix and matrix-vector operations.

%%%%%%%%%%%%%%%%%%%%%%%%%%%%%%%%%%
\begin{table}[htbp]
\begin{minipage}{.47\textwidth}
Table~\ref{tbl:synch-sgd-speedup} contains the speedup in time per iteration generated by parallel CPU over sequential CPU and GPU over parallel CPU, respectively. For synchronous SGD, this also represents the speedup in time to convergence. Given that parallel CPU uses 56 threads, we expect a speedup in the range of 56 over sequential CPU. This is the case for \texttt{news}. The speedup for \texttt{rcv1} is slightly below 56---due to the large size of the dataset which does not allow for efficient caching even in L3. We obtain super-linear speedup on \texttt{covtype}, \texttt{w8a}, and \texttt{real-sim}---on \texttt{w8a} the speedup is more than 400X for SVM. The reason for this is the improved cache behavior when all the cores are in use. \texttt{w8a} can be entirely cached in L1 due to its small size, while \texttt{real-sim} and \texttt{covtype} are cached in L2 and L3, respectively. None of these datasets can be cached on a single core when executing the sequential code. GPU improves further over parallel CPU by a factor of 1 to 5X.
\end{minipage}
\hfill
\begin{minipage}{.5\textwidth}
  \begin{center}
    \begin{tabular}{l|l||rr}

	{\textbf{task}} & {\textbf{dataset}} & {\textbf{cpu-seq/cpu-par}} & {\textbf{cpu-par/gpu}} \\

	\hline
	
	\multirow{5}{*}{LR} & covtype & 112.54 & 1.23 \\

	& w8a & 323.80 & 1.24 \\

	& real-sim & 200.46 & 2.47 \\

	& rcv1 & 46.34 & 1.52 \\

	& news & 65.27 & 5.66 \\

	\hline
	
	\multirow{5}{*}{SVM} & covtype & 99.63 & 1.32 \\

	& w8a & 428.84 & 1.03 \\

	& real-sim & 164.36 & 2.00 \\

	& rcv1 & 42.76 & 2.31 \\

	& news & 58.23 & 5.63
	
    \end{tabular}
  \end{center}
\caption{Speedup of synchronous SGD time per iteration. cpu-seq/cpu-par measures the speedup of the parallel CPU implementation over the sequential CPU. cpu-par/gpu corresponds to the speedup of GPU over parallel CPU.}\label{tbl:synch-sgd-speedup}
\end{minipage}
\hfill
\end{table}

%%%%%%%%%%%%%%%%%%%%%%%%%%%%%%%%%%
\subsubsection{Asynchronous SGD}\label{sec:experiments:results:asynch-sgd}

Based on the three dimensions of the design space and related strategies for Hogwild on GPU, we list the optimal configuration for all the datasets used in the experiments (Table~\ref{tbl:optimal-config}). The optimal configurations are determined by the fastest time to convergence for every dataset/task pair. Although \texttt{w8a} is a sparse dataset, we can afford to use a dense representation because of its relatively small dimensionality. In fact, we perform experiments with both representations and choose the better one---sparse representation turns to be better for both LR and SVM.

As expected, \texttt{covtype} benefits from coalesced col-rr data access with circular model updates on the model replicas in the shared memory---\texttt{w8a} in dense format has the same configuration. Moreover, replication is not beneficial. For the sparse datasets, the row-major format is better since it does not include zero padded features. Compared to column-major format, row-major format trades-off memory transactions by thread stalls, improving hardware efficiency considerably. Assigning examples to threads at round-robin- or chunk-level is a property of the dataset and computation due to the length of memory jumps incurred and the number of accesses per example. Due to the large model size, kernel model replication is the optimal choice for sparse datasets due to the limited size of shared memory---thread model replication degenerates into kernel with additional overhead. k-wise data replication is almost always beneficial and in larger sizes. However, the difference in time to convergence compared to no replication is small. Overall, these results confirm the importance of our study in characterizing the design and prove that applying the same configuration in all scenarios is suboptimal.

Table~\ref{tbl:asynch-sgd} depicts the time to convergence to 1\% error, and hardware and statistical efficiency for asynchronous SGD. The parallel implementation on NUMA CPU is based on DimmWitted~\cite{dimm-witted} (Section~\ref{ssec:async-sgd:cpu}). The GPU results are based on the configurations in Table~\ref{tbl:optimal-config} corresponding to each dataset/task pair.

%%%%%%%%%%%%%%%%%%%%%%%%%%%%%%%%%%
\begin{table}[htbp]
\begin{minipage}{.44\textwidth}
When comparing the CPU and GPU solutions, there is a clear trend--- \textit{(parallel) CPU is faster than GPU in time to convergence}. Specifically, on dense and low-dimensional data, the sequential CPU solution is faster, while on sparse data, parallel CPU dominates. The reason behind this Hogwild behavior on CPU is well-known~\cite{dimm-witted,hogbatch}---concurrent updates to the same features of the model generate cache-coherency conflicts that slow down execution and convergence. Essentially, parallelism is beneficial only on sparse data and models. Since there is no cache coherency mechanism on the GPU, one may expect the GPU solution to be considerably faster due to the higher degree of parallelism.
\end{minipage}
\hfill
\begin{minipage}{.56\textwidth}
  \begin{center}
    \begin{tabular}{l|l||l}
	\textbf{task} & \textbf{dataset} & \textbf{optimal configuration} \\
	\hline
	\multirow{5}{*}{LR} & covtype & col-rr + block + no-rep \\
	%& w8a-dense & col-rr + kernel + rep-10 \\
	%& sparse-w8a & row-rr + kernel + rep-5 \\
	& w8a & row-rr + kernel + rep-10 \\
	%& real-sim & row-ch + kernel + rep-2 \\
	%& real-sim &  row-ch + kernel + no-rep \\
	& real-sim & row-ch + kernel + rep-10 \\
	%& rcv1 & row-ch + kernel + rep-2 \\
	& rcv1 & row-ch + kernel + no-rep \\
	& news & row-rr + kernel + rep-10 \\
	\hline
	\multirow{5}{*}{SVM} & covtype & col-rr + block + no-rep \\
	%& w8a-dense & col-rr + kernel + rep-10 \\
	%& w8a & row-ch + kernel + no-rep \\
	& w8a & row-ch + kernel + rep-10 \\
	%& real-sim & row-ch + kernel + no-rep (for 0.7234, 51)\\
	%& real-sim & row-rr + kernel + rep-5 \\
	& real-sim & row-rr + kernel + rep-10 \\
	& rcv1 & row-rr + kernel + rep-10 \\
	%& news & row-ch + kernel + no-rep \\ (row-ch+kernel+rep-5 for 19.91,47)
	& news & row-rr + kernel + rep-10 
    \end{tabular}
  \end{center}
\caption{Optimal configurations for Hogwild on GPU.}\label{tbl:optimal-config}
\end{minipage}
\hfill
\end{table}

\noindent
However, the GPU bottleneck turns out to be vectorized execution inside a warp which generates a significant number of model update conflicts. While the warp shuffling optimization reduces the number of conflicts inside a warp, the number of concurrent warps is a lower bound that cannot be overcome. In the case of sparse data, however, update conflicts are not an issue. The problem is the irregular access to the model across the examples inside a warp. First, there is a high variance in the number of non-zero entries---several orders of magnitude. This forces threads to stall while longer examples finish. Second, all accessed model indexes have to be cached before a vectorized instruction can be executed. This incurs a large number of slow memory transactions per instruction. In order to address these issues, data and model partitioning techniques similar to the ones proposed for out-of-core processing~\cite{dot-product-join-ssdbm,hogwild-disk} have to devised for GPU. We plan to look into this topic in future work.

%%%%%%%%%%%%%%%%%%%%%%%%%%%%%%%%%%
\begin{table*}[htb]
  \begin{center}
    \begin{tabular}{l|l||rrr||rrr||rrr}

	\multirow{2}{*}{\textbf{task}} & \multirow{2}{*}{\textbf{dataset}} & \multicolumn{3}{c||}{\textbf{time to convergence (sec)}} & \multicolumn{3}{c||}{\textbf{time per iteration (msec)}} & \multicolumn{3}{c}{\textbf{\# iterations}} \\

	& & {\textbf{gpu}} & {\textbf{cpu-seq}} & {\textbf{cpu-par}} & \textbf{gpu} & \textbf{cpu-seq} & \textbf{cpu-par} & {\textbf{gpu}} & {\textbf{cpu-seq}} & {\textbf{cpu-par}} \\

	\hline
	
	\multirow{5}{*}{LR} & covtype & 1.97 & \underline{0.60} & 1.51 & \underline{15} & 150 & 251 & 135 & \underline{4} & 6 \\

	%& w8a-dense & 1.21 & 0.56 & \underline{0.47} & 0.0931(0.0098) & & 0.0207& 14($153^{\#}$) & \underline{13} & 21 \\ 

	& w8a & 0.20 & 0.27 & \underline{0.18} & 21 \underline{(2.8)} & 15 & 5.9 & \underline{8} (80) & 18 & 27 \\

	%& real-sim & 1.52$^{*}$ & 1.35 & \underline{0.52} & 19 & 25 & \underline{8.1} & 78$^{*}$ & \underline{54} & 61 \\

	& real-sim & 2.46 & 1.35 & \underline{0.52} & 271 (27) & 25 & \underline{8.1} & \underline{9} (92) & 54 & 61 \\

	& rcv1 & 18.29 & 20.37 & \underline{4.64} & 226 & 345 & \underline{71} & 81 & \underline{59} & 65 \\

	& news & 7.35 & \underline{5.47} & $\infty$ & 615 (65) & 53 & \underline{8.7} & \underline{12} ($\infty$) & 103 & $\infty$ \\

	\hline
	
	\multirow{5}{*}{SVM} & covtype & 0.96 & \underline{0.16} & 0.35 & \underline{15} & 53 & 77 & 63 & \underline{3} & 4 \\

	%& w8a-dense & 6.53 & \underline{4.53} & $2.87^{\#}$ & 0.0835(0.0096) & & 0.0196 & \underline{78}(77) & 166 & $145^{\#}$ \\

	%& w8a & \underline{0.48} & 0.56 & 1.92 & 3.3 & \underline{2.3} & 5.7 & \underline{133} & 239 & 355 \\

	& w8a & 6.29 & \underline{0.54} & 1.89 & 25 (2.6) & \underline{2.2} & 5.6 & 247 ($\infty$) & \underline{239} & 333 \\

	%& real-sim & \underline{0.58} & 1.97 & 1.24 & 53 (11) & 12 & \underline{7.6} & \underline{9} (152) & 164 & 160 \\

	& real-sim & \underline{1.04} & 1.82 & 1.28 & 136 (14) & 11 & \underline{7.6} & \underline{7} (247) & 164 & 166 \\

	& rcv1 & 8.56 & 22.71 & \underline{7.57} & 955 (94) & 216 & \underline{68} & \underline{9} (109) & 105 & 111 \\

	& news & 8.75 & 20.01 & \underline{1.79} & 454 (50) & 47 & \underline{8.4} & \underline{19} ($\infty$) & 425 & 211

    \end{tabular}
  \end{center}
\caption{Asynchronous SGD performance to 1\% convergence error. For the GPU configurations that use replication, we also include the values without replication, e.g., 21 (2.8) time per iteration for LR on \texttt{w8a} corresponds to 21 msec with rep-10 and 2.8 msec with no-rep, respectively. The best values for each dataset are underlined. $\infty$ stands for lack of convergence in 300 seconds and an unknown number of iterations, e.g., cpu-par for LR on \texttt{news} does not converge to 1\% error within 300 seconds, thus the number of iterations to convergence is unknown.}\label{tbl:asynch-sgd}
\end{table*}
%%%%%%%%%%%%%%%%%%%%%%%%%%%%%%%%%%

Several entries in Table~\ref{tbl:asynch-sgd} require discussion. 
Parallel CPU does not converge for LR on \texttt{news}---also the case for GPU without replication. This proves the benefits of accessing an example multiple times during an iteration despite the increase in hardware efficiency. Nonetheless, the decision is particular to each dataset/task pair separately---parallel CPU converges without replication in all the other configurations, while GPU does not for SVM on \texttt{w8a} and \texttt{news}.
Sequential CPU is the fastest for SVM on \texttt{w8a} because of several reasons. In sparse representation, \texttt{w8a} is small enough to be fully cached in L3, thus, memory latency is minimal. As discussed previously, parallelism triggers the cache coherency mechanism which incurs delays. Since \texttt{w8a} has only 300 features, the rate of update conflicts is high. Although this behavior is also present on LR, the results are different. This has to do with the much simpler form of the SVM gradient which can be evaluated with a simple bit flip operation---the label is +1 or -1. The LR gradient requires exponentiation which is considerably more expensive.
GPU is the fastest for SVM on \texttt{real-sim}---the only case for asynchronous SGD across all our experiments. The reason is the very small number of iterations to convergence which are executed relatively fast. This is the perfect example for data replication. Nonetheless, the gap to CPU execution is minimal.
The number of iterations sequential CPU requires to convergence for SVM on \texttt{news} is much larger than parallel CPU, e.g., 425 compared to 211. This is contrary to all the other results and unexpected. We found that sequential CPU gets stuck on a plateau for 220 iterations before the loss starts to decrease again. This is due to a too large step size that does not decrease fast enough. All our tries to find a better step size have failed---this is the best time to convergence we managed to achieve.

We report the time per iteration, i.e., hardware efficiency, and the number of iterations to convergence, i.e., statistical efficiency, for GPU with and without replication. This clearly shows the impact of replication---higher time per iteration and smaller number of iterations to convergence. While the ratio of the values follows closely the replication factor, the time to convergence is entirely determined by the actual value. Nonetheless, the times to convergence with and without replication are very close.

%%%%%%%%%%%%%%%%%%%%%%%%%%%%%%%%%%
\begin{table}[htbp]
\begin{minipage}{.47\textwidth}
In general, the best time per iteration follows closely the optimal time to convergence. Parallel CPU is the fastest on sparse data for the same reasons mentioned previously. With the exception of SVM on \texttt{w8a}, GPU has the fastest time per iteration on dense data. This proves that -- with the optimal data layout -- the superior FLOPS on the GPU lead to better performance. Identifying the optimal layout, however, requires the in-depth analysis performed in this work. The higher computational complexity of LR is reflected in the higher time per iteration across all the configurations. Table~\ref{tbl:asynch-sgd-speedup} summarizes the speedup in time per iteration corresponding to these values. In the best case, parallel CPU achieves a speedup of 6X over sequential CPU on \texttt{news} which is consistent with results published in the literature~\cite{hogbatch,cyclades}. The best speedup of GPU over parallel CPU is at most 5X on \texttt{covtype}.
\end{minipage}
\hfill
\begin{minipage}{.5\textwidth}
  \begin{center}
    \begin{tabular}{l|l||rr}

	{\textbf{task}} & {\textbf{dataset}} & {\textbf{cpu-seq/cpu-par}} & {\textbf{cpu-par/gpu}} \\

	\hline
	
	\multirow{5}{*}{LR} & covtype & 0.60 & 5.80 \\

	& w8a & 2.54 & 2.11 \\

	& real-sim & 3.09 & 0.30 \\

	& rcv1 & 4.86 & 0.31 \\

	& news & 6.09 & 0.13 \\

	\hline
	
	\multirow{5}{*}{SVM} & covtype & 0.69 & 5.13 \\

	& w8a & 0.39 & 2.15 \\

	& real-sim & 1.45 & 0.54 \\

	& rcv1 & 3.18 & 0.72 \\

	& news & 5.60 & 0.17
	
    \end{tabular}
  \end{center}
\caption{Speedup of asynchronous SGD time per iteration. cpu-seq/cpu-par measures the speedup of the parallel CPU implementation over the sequential CPU. cpu-par/gpu corresponds to the speedup of GPU over parallel CPU.}\label{tbl:asynch-sgd-speedup}
\end{minipage}
\hfill
\end{table}

\noindent
The reason it does not translate into better time to convergence is the much larger number of iterations---a factor of 15 or more. As expected, sequential CPU achieves convergence in the smallest number of iterations without data replication because it avoids update conflicts altogether. Intuitively, k-wise replication is equivalent to executing k passes over the data in a single iteration, i.e., k iterations. Asynchronous parallel processing always introduces additional iterations---correlated to the degree of parallelism.

%%%%%%%%%%%%%%%%%%%%%%%%%%%%%%%%%%
\subsubsection{CPU vs. GPU}\label{sec:experiments:results:arch}

\begin{figure*}[htbp]
\begin{center}
\includegraphics[width=\textwidth]{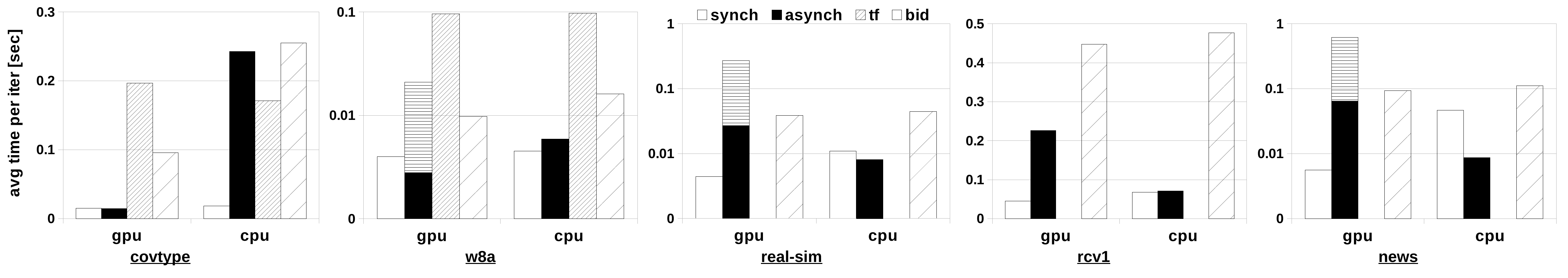}
\caption{LR hardware efficiency comparison.}\label{fig:lr-avgtime}
\includegraphics[width=\textwidth]{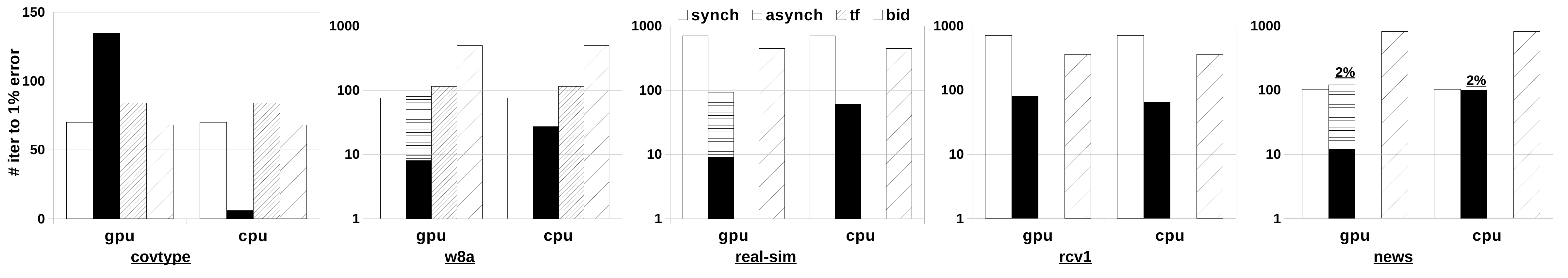}
\caption{LR statistical efficiency comparison.}\label{fig:lr-iter}
\includegraphics[width=\textwidth]{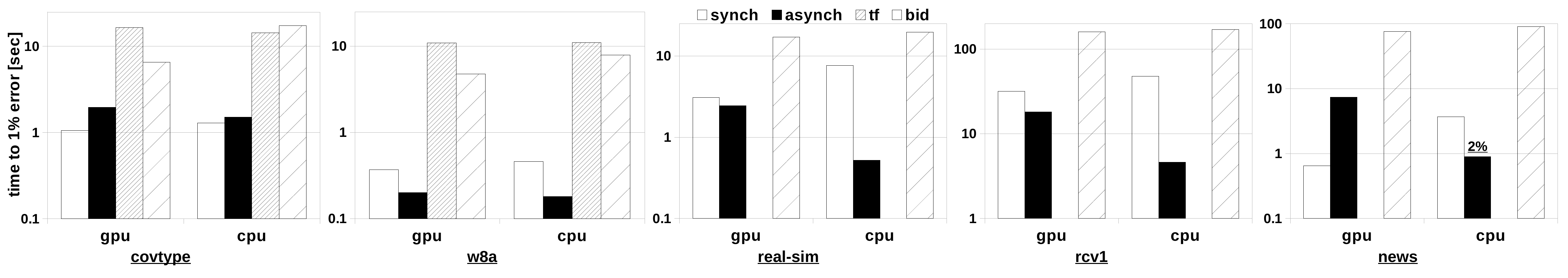}
\caption{LR time to convergence comparison.}\label{fig:lr-time}
\end{center}
\end{figure*}

\begin{figure*}[htbp]
\begin{center}
\includegraphics[width=\textwidth]{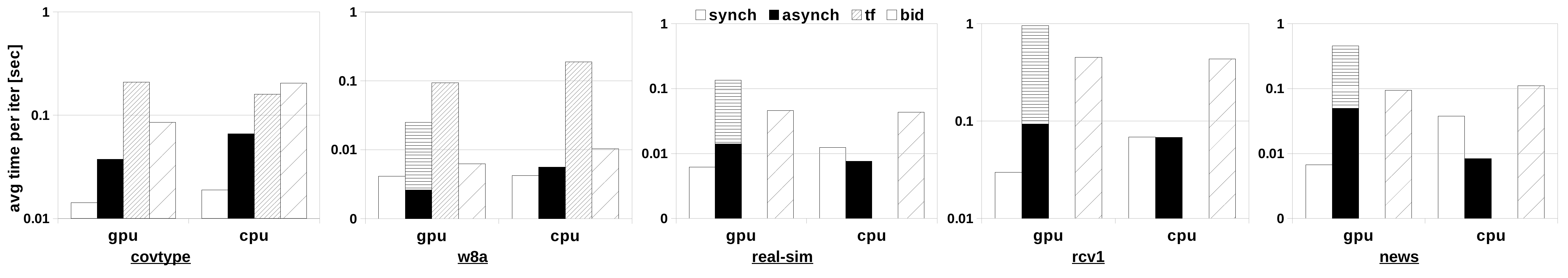}
\caption{SVM hardware efficiency comparison.}\label{fig:svm-avgtime}
\includegraphics[width=\textwidth]{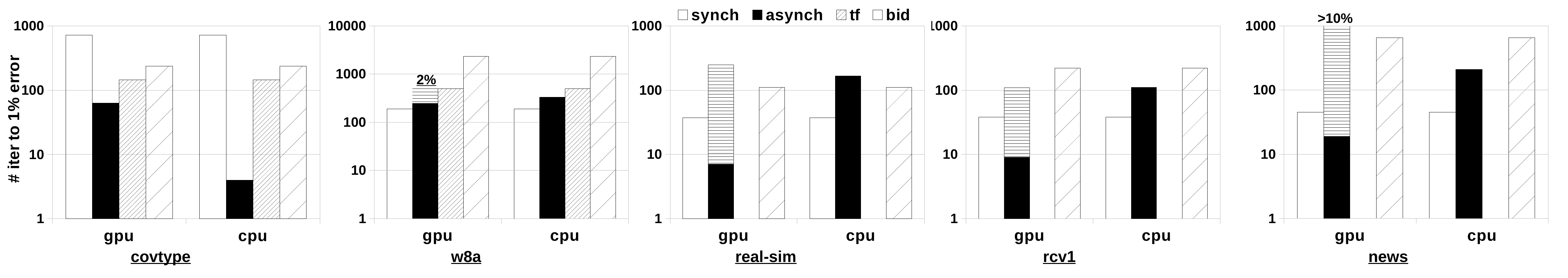}
\caption{SVM statistical efficiency comparison.}\label{fig:svm-iter}
\includegraphics[width=\textwidth]{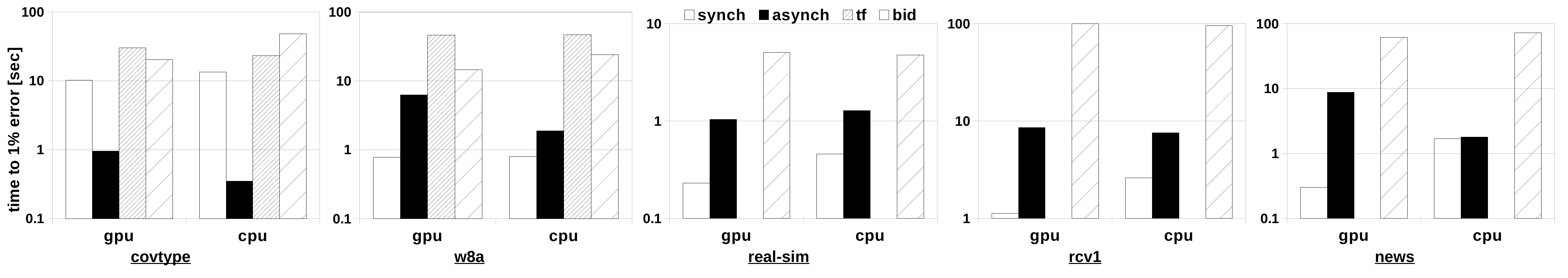}\\
\caption{SVM time to convergence comparison.}\label{fig:svm-time}
\end{center}
\end{figure*}

We group all the results -- hardware efficiency, statistical efficiency, and time to convergence -- across all the datasets in Figure~\ref{fig:lr-avgtime}--\ref{fig:lr-time} for LR and Figure~\ref{fig:svm-avgtime}--\ref{fig:svm-time} for SVM, respectively. We also include in the figures the synchronous SGD implemented in TensorFlow (tf) and BIDMach (bid)---on parallel CPU and GPU. Since TensorFlow supports only dense data, we have results only for \texttt{covtype} and \texttt{w8a}. The stacked bars for asynchronous SGD on GPU correspond to the configurations with and without replication---hardware efficiency is smaller without replication, while statistical efficiency is smaller with replication. The side-by-side depiction of the results allows for immediate comparison between CPU and GPU across four different implementations. It also allows us to compare our ViennaCL synchronous SGD with the ones implemented in TensorFlow and BIDMach.

%%%%%%%%%%%%%%%%%%%%%%%%%%%%%%%%%%%%
%%%%%%%%%%%%%%%%%%%%%%%%%%%%%%%%%%%%%

For our parallel implementations -- synch and asynch -- the hardware efficiency results in Figure~\ref{fig:lr-avgtime} and~\ref{fig:svm-avgtime} are another representation of the time per iteration results from Table~\ref{tbl:synch-bgd} and~\ref{tbl:asynch-sgd}. As discussed previously, our synchronous GPU always outperforms synchronous CPU because of the higher degree of parallelism. We observe the same pattern for synchronous SGD in BIDMach---maybe with a smaller gap on sparse data. The TensorFlow results, while limited to dense data, are somehow different. Parallel CPU is almost identical and even better than GPU---on \texttt{covtype}. We found matrix transpose computation to be the bottleneck on the GPU---the same problem as in ViennaCL. The hardware efficiency of asynchronous SGD is more sensitive to the task and dataset. Nonetheless, the trend in these figures suggests that GPU is better on dense data, while parallel CPU is better for sparse data. The reasons are discussed in Section~\ref{sec:experiments:results:asynch-sgd}. In terms of statistical efficiency, all the synchronous SGD implementations -- including TensorFlow and BIDMach -- require exactly the same number of iterations to converge to a given loss accuracy both on CPU and GPU. While the results confirm this, they also show that statistical efficiency is sensitive to the task and dataset. The statistical efficiency of asynchronous SGD on CPU is always better than on GPU -- without replication -- because of the reduced number of model update conflicts generated by a smaller number of threads. The gap between the two is especially high on dense data and small models.

The time to convergence for synchronous SGD is a scaled-up image of the time per iteration since the number of iterations is identical on CPU and GPU. Thus, with the exception of TensorFlow, synchronous GPU always converges faster than CPU. In addition to the matrix transpose inefficiency, the GPU kernels in TensorFlow are optimized for dense matrices appearing in deep nets convolutions. These operations are more compute-intensive than the matrix-vector multiplications in LR and SVM. Since BIDMach is not exclusively targeted at deep learning, it optimizes these kernels better, while our implementation is focused on generalized linear models. The time to convergence for asynchronous SGD is a direct rendering of the results in Table~\ref{tbl:asynch-sgd}. For the reasons discussed in that context, we observe that the CPU implementation always outperforms GPU, even though GPU has better hardware efficiency on dense data. However, this is not sufficient to compensate for the much higher number of iterations to converge.

%%%%%%%%%%%%%%%%%%%%%%%%%%%%%%%%%%%%
%%%%%%%%%%%%%%%%%%%%%%%%%%%%%%%%%%%%%

%%%%%%%%%%%%%%%%%%%%%%%%%%%%%%%%%%
\subsubsection{Synchronous vs. asynchronous}\label{sec:experiments:results:sync-async}

Figure~\ref{fig:lr-avgtime}--\ref{fig:lr-time} and Figure~\ref{fig:svm-avgtime}--\ref{fig:svm-time} also provide a comparison between synchronous and asynchronous SGD for CPU and GPU, respectively. Since these are different algorithms initialized with different hyper-parameters, the only comparison that makes sense is in terms of hardware efficiency---the statistical efficiency and, thus, the time to convergence, are properties of the task and dataset. On the GPU, the hardware efficiency of synchronous SGD is generally better than that of asynchronous SGD. On the CPU, the hardware efficiency of synchronous SGD is better only for low-dimensional models, while for sparse high-dimensional data asynchronous is slightly better. To understand these results, we have to discuss first the two implementations. Synchronous SGD consists of a series of simple kernels -- one for each linear algebra primitive -- that process the complete input data. The result of each kernel is fully-materialized in memory. Essentially, synchronous SGD has a materialization execution strategy. Asynchronous SGD -- on the other hand -- consists of a single kernel that fuses all the gradient operations. Moreover, it also updates the model for each example, thus, executes more operations. In view of these, materialization is preferred on the GPU -- as long as it does not incur memory overflow -- because of simpler memory access patterns. The complete elimination of model updates -- equal to the number of examples -- is also an important factor. On the CPU, materialization is bounded by the size of the caches, not the complete memory---the smaller the intermediate results the better. Since none of the datasets generates small-enough intermediates that can be cached in the upper layers of the hierarchy, a large number of cache misses that degrade performance is incurred. While operator fusion -- or compilation -- improves cache access, the cache coherency mechanism triggered by concurrent model updates continues to be an important deterrent for performance. On low-dimensional models, fusion cannot overcome the large number of model update conflicts per example. On sparse high-dimensional models with large intermediates, however, the number of conflicts is minor. Thus, asynchronous SGD outperforms synchronous SGD in hardware efficiency.

%%%%%%%%%%%%%%%%%%%%%%%%%%%%%%%%%%%%
\begin{figure*}[htbp]
\begin{center}
\includegraphics[width=\textwidth]{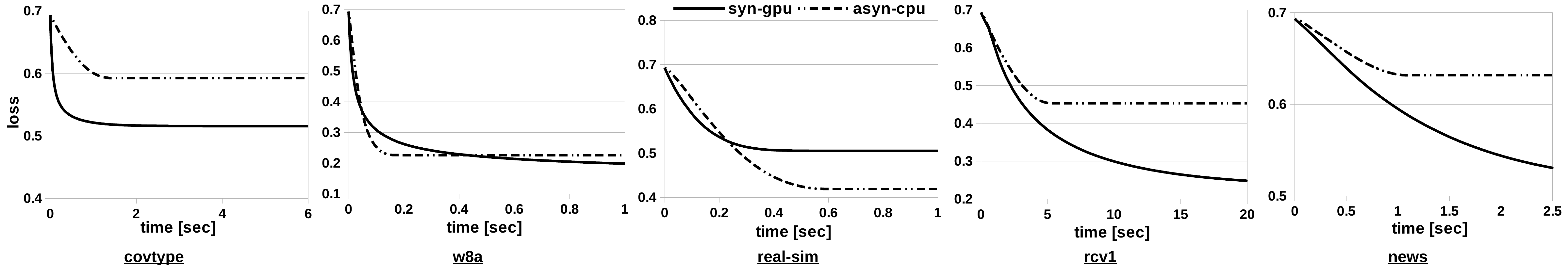}
\caption{Comparison in time to convergence between synchronous GPU and asynchronous CPU on LR.}\label{fig:lr-loss}
\includegraphics[width=\textwidth]{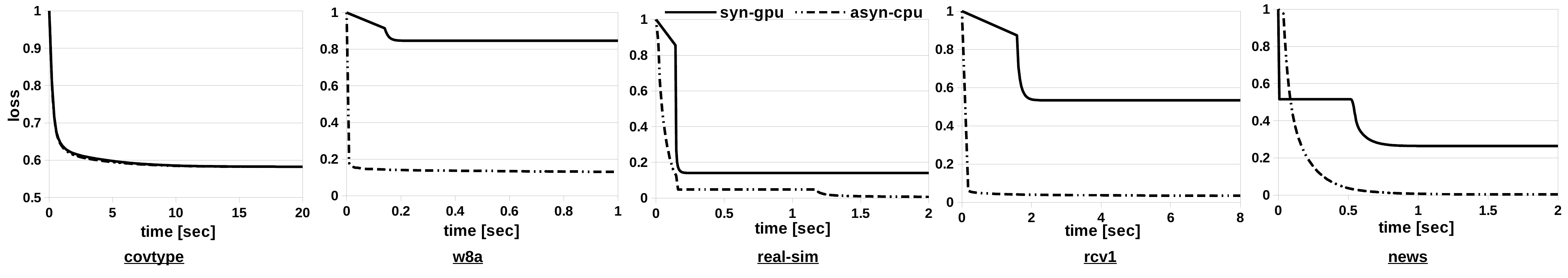}
\caption{Comparison in time to convergence between synchronous GPU and asynchronous CPU on SVM.}\label{fig:svm-loss}
\end{center}
\end{figure*}
%%%%%%%%%%%%%%%%%%%%%%%%%%%%%%%%%%%%%

We perform a direct comparison in time to convergence only between synchronous GPU and asynchronous CPU---the optimal configurations identified for each model update strategy. We measure the loss as a function of time for exactly the same hyper-parameters and the same initialization conditions. This allows us to isolate the effect of the update strategy while using the optimal computing architecture. The results are depicted in Figure~\ref{fig:lr-loss} for LR and Figure~\ref{fig:svm-loss} for SVM, respectively. Synchronous GPU achieves better convergence for certain dataset/task pairs, while asynchronous CPU is better for others. Specifically, synchronous GPU dominates on LR, while asynchronous CPU dominates on SVM. Given that this is essentially a comparison between batch gradient descent -- which corresponds to synchronous GPU -- and stochastic gradient descent -- which corresponds to asynchronous CPU -- we do not expect a single winner all the time. As shown previously in the literature~\cite{gd-optimization}, the best optimization strategy is particular to the task and the dataset. Our results confirm this finding for parallel optimizers with different model update strategies---a new scenario that has not been studied before. To summarize, \textit{while GPU is the optimal architecture for synchronous SGD and CPU is optimal for asynchronous SGD, choosing the better of synchronous GPU and asynchronous CPU is task- and dataset-dependent}.

%%%%%%%%%%%%%%%%%%%%%%%%%%%%%%%%%%
\subsubsection{Comparison with TensorFlow and BIDMach}\label{sec:experiments:results:tf-bid}

We compare our synchronous SGD implementation in ViennaCL with the solutions in TensorFlow and BIDMach based on the results in Figure~\ref{fig:lr-avgtime}--\ref{fig:lr-time} and Figure~\ref{fig:svm-avgtime}--\ref{fig:svm-time}. The main point of this comparison is only to verify that our implementation is efficient. We observe that our synchronous SGD outperforms both TensorFlow and BIDMach in time per iteration and time to convergence for all the datasets and all the tasks---both on CPU and GPU. We emphasize that we measure only the time spent in critical processing across all the solutions. The performance of TensorFlow and BIDMach on dense data is comparable. TensorFlow is slightly better on the CPU because of the Java/Scala overhead incurred by BIDMach---which is better on GPU. As previously discussed, TensorFlow has an inefficient matrix transpose on GPU. In terms of statistical efficiency, there are cases where our synchronous SGD requires more iterations to converge. Nonetheless, they are rare and when they occur, the statistical efficiency of our asynchronous SGD is better than that of TensorFlow and BIDMach. The reason there are differences between the synchronous implementations despite them being algorithmically identical is different linear algebra kernels and different model update protocols. In summary, \textit{our SGD implementations always outperform TensorFlow and BIDMach in time per iteration, number of iterations to convergence, and time to convergence}.

%%%%%%%%%%%%%%%%%%%%%%%%%%%%%%%%%%%%
\begin{figure*}[htbp]
\begin{center}
\includegraphics[width=\textwidth]{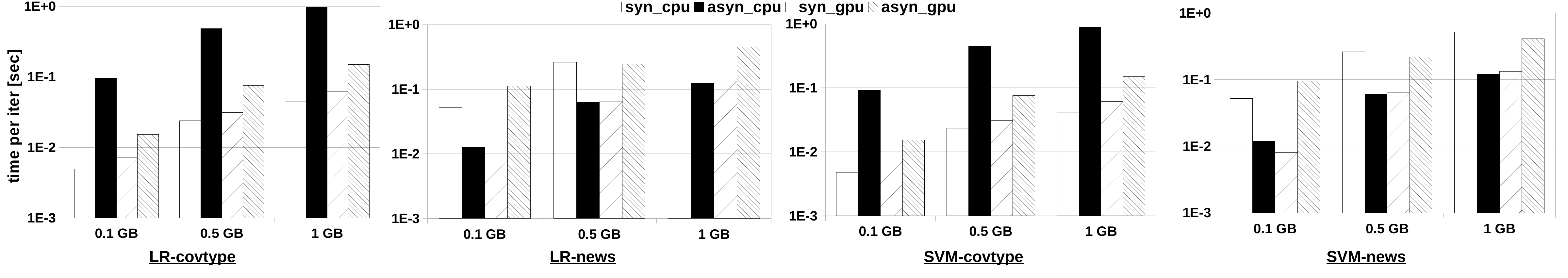}
\caption{Hardware efficiency as a function of the number of examples or size of the taining dataset.}\label{fig:scal-lr-svm}
\end{center}
\end{figure*}
%%%%%%%%%%%%%%%%%%%%%%%%%%%%%%%%%%%%%

%%%%%%%%%%%%%%%%%%%%%%%%%%%%%%%%%%
\subsubsection{Scalability with the number of examples}\label{sec:experiments:results:scal-examples}

In order to study the effect of the number of examples in the training dataset -- or the size of the dataset -- on the hardware efficiency of our algorithms, we generate three different instances of the \texttt{covtype} and \texttt{news} datasets with increasing number of examples. The size of these datasets is 100 MB, 500 MB, and 1 GB, respectively. We execute all the algorithms on these datasets and measure the time per iteration. The results are depicted in Figure~\ref{fig:scal-lr-svm}. As expected, the time per iteration increases almost linearly with the increase of the dataset size. More importantly, the relative ordering of the algorithms is almost always preserved. Inversions happen on the sparse \texttt{news} dataset and they are minor. Specifically, asynchronous GPU becomes slightly faster than synchronous CPU, while asynchronous CPU becomes slightly faster than synchronous GPU. To put it differently, the synchronous solutions on sparse data are impacted negatively by larger training datasets. The reason is the increase in size of the intermediate results which deteriorate cache efficiency on CPU and memory usage on GPU, respectively. The rate of model update conflicts on dense data is too high for this behavior to be observed. We point out that we cannot study the effect of dataset size on convergence because the value of the loss function changes with the number of examples.

%%%%%%%%%%%%%%%%%%%%%%%%%%%%%%%%%%%%
\begin{figure*}[htbp]
\begin{center}
\includegraphics[width=\textwidth]{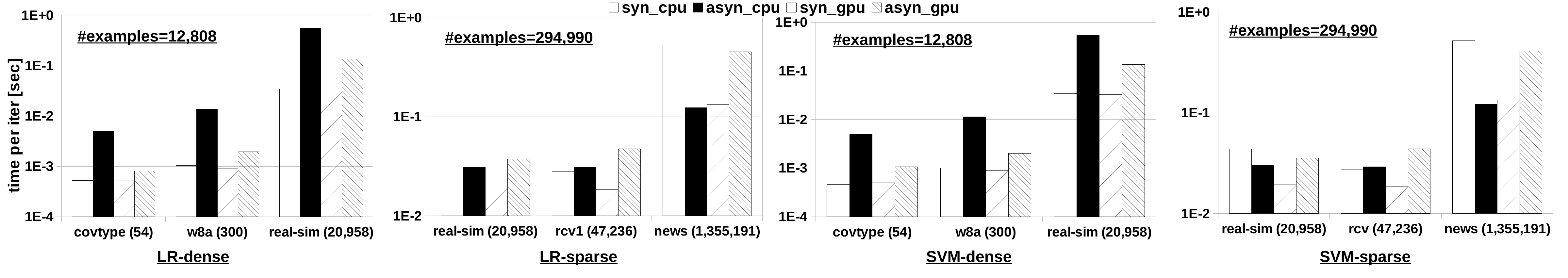}
\caption{Hardware efficiency as a function of the number of features or dimensions in dense and sparse format.}\label{fig:dim-lr-svm}
\end{center}
\end{figure*}
%%%%%%%%%%%%%%%%%%%%%%%%%%%%%%%%%%%%%

%%%%%%%%%%%%%%%%%%%%%%%%%%%%%%%%%%
\subsubsection{Scalability with the number of features}\label{sec:experiments:results:scal-features}

We take the three lowest-dimensional datasets -- \texttt{covtype}, \texttt{w8a}, and \texttt{real-sim} -- and materialize them in dense format while keeping the number of examples identical. The largest generated dataset, i.e., \texttt{real-sim}, is 2.15 GB in size. In order to generate data in sparse format, we extract the same number of examples from the three highest-dimensional datasets---\texttt{real-sim}, \texttt{rcv1}, and \texttt{news}. With this number of examples, \texttt{news} has the largest size---approximately 2 GB. We execute all the proposed algorithms on all the datasets for LR and SVM. Figure~\ref{fig:dim-lr-svm} depicts hardware efficiency as a function of the number of features. On data in dense format -- as expected -- the time per iteration increases with the increase in dimensionality. Unlike the increase with the number of examples, in this case we have a slightly sub-linear increase that impacts all the configurations equally---the relative order of the algorithms is preserved with the increase in dimensionality. For synchronous algorithms, the increase is due entirely to handling larger matrices. In the case of asynchronous algorithms, the examples become complete through ``densification'', thus, they access the entire model---the semantics of ``zero'' is unknown during execution. As a result, the time per iteration is proportional to the number of features. The results on sparse data are more intriguing. First, we observe that going from \texttt{real-sim} to \texttt{rcv1} -- doubling the number of features -- reduces the time per iteration for the synchronous algorithms. The increase in dimensionality, however, comes with a relatively small increase in the average number of non-zero dimensions per example. Moreover, the examples in \texttt{rcv1} display less variance in the number of non-zeros compared to \texttt{real-sim}---\texttt{rcv1} is more homogeneous than \texttt{real-sim}. We observe that ViennaCL handles the more uniform \texttt{rcv1} matrix more efficiently with respect to cache and memory access, thus, the reduced time per iteration. The other phenomenon we observe on sparse data is that the relative performance of asynchronous CPU and synchronous GPU switches with the increase in dimensionality. This is similar to the behavior in Figure~\ref{fig:scal-lr-svm} and the reasons are the same---\textit{asynchronous CPU outperforms synchronous GPU on highly-dimensional models trained over a sufficiently large number of examples}.

%%%%%%%%%%%%%%%%%%%%%%%%%%%%%%%%%%
\subsection{Summary}\label{sec:experiments:summary}

Based on the extensive experiments we perform, we are in the position to provide answers to the questions identified at the beginning of the section. We repeat the questions -- this time with the corresponding answers -- in the following:
\begin{compactitem}
\item What is role of the computing architecture, i.e., CPU/GPU, on the performance of synchronous SGD? \textit{GPU always outperforms parallel CPU in hardware efficiency and, consequently, in time to convergence. The difference is minimal for small low-dimensional datasets and increases with dimensionality and sparsity---for a maximum speedup of 5.66X.}
\item What is role of the computing architecture, i.e., CPU/GPU, on the performance of asynchronous SGD? \textit{Although parallel CPU outperforms GPU in general, it is more difficult to identify the optimal computing architecture for asynchronous SGD. The main reason is the complex interaction between hardware and statistical efficiency.}
\item What is the optimal configuration for asynchronous SGD on GPU? \textit{The optimal data access path + model replication + data replication configuration depends on the task and the training dataset. Limiting any implementation to a single configuration results in sub-optimal -- and sometimes no -- convergence.}
\item How do synchronous and asynchronous SGD compare against each other on CPU and GPU separately, and across computing platforms? \textit{Synchronous SGD is the optimal choice on GPU and asynchronous SGD is the safe choice on CPU. The better choice between these two depends on the task and the training dataset since they mirror the comparison between BGD and SGD.}
\item Are our implementations efficient with respect to TensroFlow and BIDMach? \textit{Our synchronous SGD -- which is the equivalent of the TensorFlow and BIDMach implementations -- is always faster in time to convergence both on CPU and GPU.}
\item How do the proposed algorithms scale with the number of training examples and the dimensionality of the feature vector, respectively? \textit{On dense data, the increase in time per iteration is proportional to the increase in data size for all the algorithms. On sparse data, the distribution of the non-zero entries impact scalability to a similar degree as model dimensionality---in certain cases even more.}
\end{compactitem}

%%%%%%%%%%%%%%%%%%%%%%%%%%%%%%%%%%%%%%%%%%%%%%%%%%%%%%%%%%%%%%%%%
%\input{rel-work}

\section{RELATED WORK}\label{sec:rel-work}

%%%%%%%%%%%%%%%%%%%%%%%%%%%%%%%%%%
\textbf{SGD on CPU.}
SGD is the most popular optimization method to train analytics models. Bismarck~\cite{bismarck} and GLADE~\cite{igd-glade} present methods to implement SGD inside a database engine. DimmWitted~\cite{dimm-witted} provides a study on how to implement parallel SGD on NUMA architectures. While similar exploratory axes and measure terminology are introduced, the focus on GPU is what distinguishes our paper from DimmWitted. Hogwild~\cite{hogwild} performs model updates concurrently and asynchronously without locks. Due to this simplicity -- and the near-linear speedup -- Hogwild is widely used in many analytics tasks~\cite{RRTB12,bismarck,LWR+14,DJM13,google-brain,project-adam}. Hogbatch~\cite{hogbatch} is an extension to Hogwild that is more scalable to cache-coherent architectures, while Cyclades~\cite{cyclades} reduces model update conflicts using graph partitioning. Hogwild extensions to big models based on model partitioning are introduced in~\cite{dot-product-join-ssdbm,hogwild-disk}. Buckwild~\cite{buckwild} is a low-precision variant of Hogwild that represents the data and model with fewer bits. Model averaging~\cite{parallel-igd} is an alternative method to parallelize SGD that is adequate in distributed settings. A detailed experimental comparison of Hogwild and averaging is provided in~\cite{igd-glade-ola}. The integration of relational join with gradient computation has been studied in~\cite{sgd-over-join,sgd-over-join-2,bgd-over-factorized-join}. These solutions work only for batch gradient descent (BGD), not SGD. A cost-based optimizer that selects between sequential BGD and SGD is proposed in~\cite{gd-optimization}.

%%%%%%%%%%%%%%%%%%%%%%%%%%%%%%%%%%
\textbf{SGD on GPU.}
SGD is supported by all the major deep learning frameworks, including Caffe, TensorFlow, MXNet, BIDMach, SINGA, Theano, and Torch. These frameworks implement optimized kernels for GPU processing. As far as we can tell, all these kernels are for synchronous SGD---there is no Hogwild GPU kernel. As pointed out in~\cite{Caffe-con-Troll}, since convolutions are the most expensive operation in deep learning, they are the main candidate for offloading on GPU. GeePS~\cite{geeps} implements a distributed parameter server for training across multiple GPUs. Omnivore~\cite{Omnivore} is an optimizer for deep learning on CPU and GPU that achieves better SGD performance because of careful data partitioning and placement. The asynchronous SGD supported in Omnivore is cross-device, not within the GPU---the case in our work. GPUs are effectively used for querying deep neural networks in NoScope~\cite{CuMF-SGD}. The work outside deep learning is targeting low-rank matrix factorization for recommender systems. In~\cite{MF-SGD-GPU}, dynamic scheduling strategies for low-rank matrix factorization on GPU are explored. The problem is modeled as a graph and scheduling is executed for independent subgraphs which do not have update conflicts. cuMF\_SGD~\cite{CuMF-SGD} extends dynamic scheduling with optimized SGD kernels that leverage the GPU cache, warp-shuffle instructions, and low-precision arithmetic. This is the only Hogwild GPU kernel we found in the literature. However, the design space is not explored at all.

%%%%%%%%%%%%%%%%%%%%%%%%%%%%%%%%%%%%%%%%%%%%%%%%%%%%%%%%%%%%%%%%%
%\input{conclusions}

\section{CONCLUSIONS AND FUTURE WORK}\label{sec:conclusions}

In this paper, we perform a comprehensive study of parallel SGD for generalized linear models over NUMA CPU and GPU architectures. We measure hardware efficiency, statistical efficiency, and time to convergence as a function of the objective function, model updates, and data sparsity. Overall, our study shows that the optimal SGD solution for a given architecture is highly-dependent on all of these factors. Thus, the main value of this work is to map the overall solution space and provide a useful guide for applying parallel SGD in practice. In the process, we also design several optimizations for asynchronous SGD on GPU which have their stand-alone value. We draw several interesting insights from our extensive experimental study on five real datasets. For synchronous SGD, GPU always outperforms parallel CPU---they both outperform a sequential CPU solution by more than 400X. For asynchronous SGD, parallel CPU is the safest choice while GPU with data replication is better in certain situations. The choice between synchronous GPU and asynchronous CPU depends on the task and the characteristics of the data. While LR and SVM are wide-spread ML tasks, it is intriguing to see how the results extend to other types of models, such as low-rank matrix factorization and deep neural nets. This is a topic we will pursue in future work. We point out that in low-rank matrix factorization the structure of the problem imposes limitations that are not SGD-specific, but rather method-specific. Similarly, in convolution deep nets, computing the convolution kernels is the most time-consuming operation. By focusing on simpler tasks such as LR and SVM, we are able to quantify the exclusive performance of the SGD algorithms. In the future, we also plan to consider low-precision formats in data representation and study heterogeneous solutions that integrate concurrent processing across the CPU and GPU. We also intend to explore optimization strategies that select between synchronous GPU and asynchronous CPU dynamically.

%%%%%%%%%%%%%%%%%%%%%%%%%%%%%%%%%%%%%%%%%%%%%%%%%%%%%%%
\paragraph*{Acknowledgments}
This work is supported by a U.S. Department of Energy Early Career Award (DOE Career).

%%%%%%%%%%%%%%%%%%%%%%%%%%%%%%%%%%%%%%%%%%%%%%%%%%%%%%%%%%%%%%%%%
\bibliographystyle{abbrv}
%\bibliography{biblio}

%%%%%%%%%%%%%%%%%%%%%%%%%%%%%%%%%%%%%%%%%%%%%%%%%%%%%%%%%%%%%%%%%

\end{document}